\documentclass[aps,pre,twocolumn]{revtex4}

\usepackage{epsfig}

\usepackage{graphicx}

\newcommand {\be} {\begin{equation}}
\newcommand {\ee} {\end{equation}}
\newcommand {\bea} {\begin{eqnarray}}
\newcommand {\eea} {\end{eqnarray}}
\newcommand {\bes} {\begin{displaymath}}
\newcommand {\ees} {\end{displaymath}}
\newcommand {\beas} {\begin{eqnarray*}}
\newcommand {\eeas} {\end{eqnarray*}}

\begin{document}

\title{Diffusion of Finite-Sized Hard-Core Interacting Particles In a One-Dimensional Box - Tagged Particle Dynamics}

\author{L.~Lizana}
\email{lizana@nbi.dk}
\affiliation{Niels Bohr Institute, Blegdamsvej 17, DK-2100 Copenhagen, Denmark }

\author{T.~Ambj\"ornsson}
\email{tobias.ambjornsson@thep.lu.se}
\affiliation{ Department of Theoretical Physics, Lund University, S\"olvegatan 14A, SE-223 62 Lund, Sweden,}
\affiliation{Department of Chemistry, Massachusetts Institute of
  Technology. 77 Massachusetts Avenue, Cambridge, MA 02139, USA.}

\date{\today}



\begin{abstract}
  We solve a non-equilibrium statistical mechanics problem exactly,
  namely, the single-file dynamics of $N$ hard-core interacting
  particles (the particles cannot pass each other) of size $\Delta$
  diffusing in a one dimensional system of finite length $L$ with
  reflecting boundaries at the ends.  We obtain an exact expression
  for the conditional probability density function $\rho_{\cal
    T}(y_{\cal T},t|y_{{\cal T},0})$ that a tagged particle ${\cal T}$
  (${\cal T}=1,...,N$) is at position $y_{\cal T}$ at time $t$ given
  that it at time $t=0$ was at position $y_{{\cal T},0}$.  Using a
  Bethe ansatz we obtain the $N$-particle probability density
  function, and by integrating out the coordinates (and averaging over
  initial positions) of all particles but particle ${\cal T}$, we
  arrive at an exact expression for $\rho_{\cal T}(y_{\cal T},t|y_{{\cal
      T},0})$ in terms of Jacobi polynomials or hypergeometric
  functions. Going beyond previous studies, we consider the asymptotic
  limit of large $N$, maintaining $L$ {\em finite}, using a
  non-standard asymptotic technique. We derive an exact expression for
  $\rho_{\cal T}(y_{\cal T},t|y_{{\cal T},0})$ for a  tagged particle
  located roughly in the middle of the system, from which we find that
  there are three time regimes of interest for finite-sized systems:
  $(A)$ For times much smaller than the collision time $t\ll \tau_{\rm
    coll}=1/(\varrho^2D)$, where $\varrho=N/L$ is the particle
  concentration and $D$ the diffusion constant for each particle, the
  tagged particle undergoes normal diffusion; $(B)$ for times much
  larger than the collision time $ t\gg \tau_{\rm coll}$ but times
  smaller than the equilibrium time $t\ll \tau_{\rm eq}=L^2/D$ we find
  a single-file regime where $\rho_{\cal T}(y_{\cal T},t|y_{{\cal
      T},0})$ is a Gaussian with a mean square displacement scaling as
  $t^{1/2}$; $(C)$ For times longer than the equilibrium time $t\gg
  \tau_{\rm eq}$, $\rho_{\cal T}(y_{\cal T},t|y_{{\cal T},0})$
  approaches a polynomial-type equilibrium probability density
  function.  Notably, only regimes $(A)$ and $(B)$ are found in the
  previously considered infinite systems.

\end{abstract}

\maketitle


\section{Introduction} \label{sec:Introduction}

Recent development of single fluorophore tracking techniques allow
experimental studies of the motion of particles in cellular environments with
nanometer resolution \cite{Rust_06}. The cell interior represents a crowded
environment, in which the motion of an individual particle is strongly
effected by the presence of other particles. Crowding affects, for instance,
the folding of proteins, diffusional motion \cite{LBE,WAC} as well as rates of
biochemical reactions \cite{Luby_Phelps_00, Ellis_03}. Crowding is also
important during ribosomal translation on mRNA \cite{MSP}, and binding protein
diffusion along DNA \cite{Halford_04, Lomholt_05}, where bound proteins are
hindered from passing each other. Furthermore, advances in nanofluidics allow
studies of geometrically constrained nano-sized particles
\cite{Dekker_07,Karlsson_01}.  The system considered in this paper, the
diffusion of a tagged particle immersed in a one-dimensional bath of hard-core
interacting particles - in the literature referred to as single-file diffusion
(SFD) - represents one of the simplest systems governed by crowding effects,
but yet with possible applications for obstructed one-dimensional protein
diffusion along DNA molecules and transport in nanofluidic systems.

SFD phenomena emerge in quasi one-dimensional geometries. The particle
order is under these circumstances conserved over time $t$ which
results in interesting dynamics for a tagged particle, quite different
from what is predicted from classical diffusion (governed by Fick's
law).  Examples found in nature include ion or water transport through
pores in biological membranes~\cite{HOKE}, one-dimensional hopping
conductivity~\cite{MR} and channeling in zeolites~\cite{KKDGPRSUK}.
SFD effects have also been studied in a number of experimental setups
such as colloidal systems and ring-like
constructions~\cite{WBL,CJG,CJG2,LKB,Lin_05}. One of the most apparent
characteristics of SFD is that the mean square displacement (MSD)
${\cal S}(t) = \langle (y_{\cal T}-y_{{\cal T},0} )^2\rangle $ of a
tagged particle is in the long time limit proportional to $t^{1/2}$
in an infinite system with a fixed particle concentration (brackets
denote an average over initial positions and noise, $y_{\cal T}$ and
$y_{{\cal T},0}$ are tagged particle positions at time $t$ and $t=0$,
respectively).  Also, the conditional probability density function for
the tagged particle position (tPDF) is Gaussian.

The first theoretical study showing the $t^{1/2}$ law of the MSD and
that the tPDF is Gaussian is Ref. \cite{HA}. Subsequent studies,
proving the MSD law in alternative ways, are found in
\cite{LE,Beijeren_83,Hahn_95,Arratia_83}. Simple arguments to its
origin are presented in \cite{Percus_74,Karger_93,
  Alexander_Pincus_78}, one of which \cite{Alexander_Pincus_78} uses a
simple relationship between the displacement of a single particle and
 particle density fluctuations, the latter known to be the same as for
independent particles \cite{HA,Kutner_81}.
%
%
The $t^{1/2}$-law and Gaussian behavior have, in the long time limit,
been shown to be of general validity for identical strongly
over-damped particles interacting via any short range potential in
which mutual passage is forbidden \cite{Kollmann_03}. A generalized
central limit theorem for tagged particle motion has also been proven
\cite{Jara_Landim_06}.
More recent work include \cite{Nelissen_07} where particles
interacting via a screened Coulomb potential (i.e. not perfectly
hard-core) was studied numerically (see also \cite{semihardcore}),
\cite{eli_prl} deals with SFD in an external potential, and
\cite{goncalves,diff_D,aslangul_random_D} address SFD dynamics with
different diffusion constants. A phenomenological Langevin formulation
of SFD was presented in \cite{TAMA}.

Although much work has been dedicated to single-file systems, to our
knowledge, very few exact results for {\rm finite} systems with
finite-sized particles have been obtained. The one exception is
\cite{RKH} where the PDF for $N$ point-particles diffusing on a finite
one-dimensional line was derived. However, simplified expressions for
the tPDF was only considered in the thermodynamic limit
($N,L\rightarrow \infty$ where $L$ denotes system length and the
concentration $\varrho=N/L$ kept fixed). In this paper, we go beyond
previous studies in the following ways: First, finite-sized particles
are considered and we show that the $N$-particle probability density
function ($N$PDF) can be written as a Bethe ansatz solution. We obtain
an exact expression for the tPDF in terms of Jacobi polynomials (or
hypergeometric functions) which reduces to that in \cite{RKH} for the
case of point particles. Second, we perform a (non-standard) large
$N$-analysis of the tPDF, keeping the system size $L$ {\em finite}.
The expression for $\rho_{\cal T}(y_{\cal T},t|y_{{\cal T},0})$ in the
many-particle limit is new and is presented compactly in terms of
modified Bessel functions. An analysis of the tPDF reveals the
existence of three dynamical regimes for a particle located roughly in
the middle of the system: $(A)$ short times, $t \ll \tau_{\rm
  coll}=1/(\varrho^2D)$ where $\tau_{\rm coll}$ denotes the collision
time and $D$ is the diffusion constant. In this limit the tagged
particle undergoes standard Brownian motion with a MSD ${\cal S}(t)
\sim t$; $(B)$ For intermediate times, $\tau_{\rm coll} \ll t \ll
\tau_{\rm eq}$ where $\tau_{\rm eq}=L^2/D$ is the equilibrium time, we
get a SFD regime where ${\cal S}(t) \sim t^{1/2}$. $(C)$ for long
times, $t\gg \tau_{\rm eq}$, an equilibrium tPDF of polynomial-type is
found. Notably, only regimes $(A)$ and $(B)$ exist in infinite
systems.

This paper has the following organization:
Sec. \ref{sec:ProblemDef} contains the formulation of the problem and a mapping onto a point particle system. The tPDF is
also formally stated in terms of the $N$PDF to which governing
dynamical equations are introduced.
In Sec. \ref{sec:NpartDist}, we provide the solution to the equations
of motion for the $N$PDF using a coordinate Bethe ansatz.
In Sec. \ref{sec:P_bar_general} the initial coordinates as well as the
coordinates for all particles except the tagged one are integrated out
in order to obtain an exact expression for the tPDF. Also, asymptotic
results for large $N$ for the tPDF are derived.
In Sec. \ref{sec:three_regimes} the asymptotic large $N$ expression
for the tPDF is expanded for short and long times, and three different
time regimes $(A)-(C)$ (see above) are identified.
More technical details are given in the appendices. A brief
summary of some of our results, corroborated with Gillespie
simulations (Monte Carlo-type), can be found in~\cite{our_prl}.


\section{Problem definition}\label{sec:ProblemDef}

\begin{figure}
 \includegraphics[width=0.9\columnwidth]{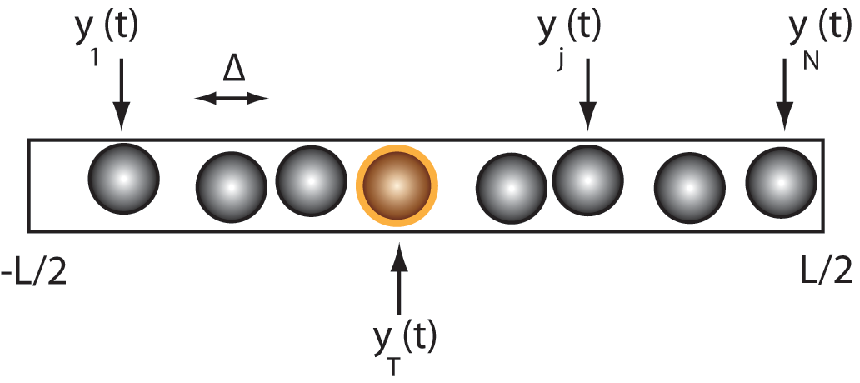} 
 \caption{(color online) Cartoon of the problem considered here: $N$
   particles of linear size $\Delta$ diffusing is a one-dimensional
   system of length $L$. The particles have center-of-mass coordinates $y_j$ and
   initial positions $y_{j,0}$, $j=1,..,N$, and are unable to overtake.
   This implies that $y_{j+1}\geq y_j+\Delta$ ($j=1,..,N-1$) for all times. 
   Also, the
   particles cannot diffuse out of the box, i.e. $y_1>-L/2+\Delta/2$
   and $y_N<L/2-\Delta/2$. }
 \label{fig:cartoon}
\end{figure}

In this paper we consider a system of $N$ identical hard-core
interacting particles, each with diffusion a constant $D$ and a linear
size $\Delta$, diffusing in a finite one dimensional system extending
from $-L/2$ to $L/2$.  A schematic cartoon is depicted in
Fig.~\ref{fig:cartoon}. The particles each have center-of-mass and
initial coordinates $\vec{y}=(y_1,\ldots,y_N)$ and
$\vec{y}_0=(y_{1,0},\ldots,y_{N,0})$, respectively. Due to the
hard-core interaction, the particles cannot pass each other and retain
their order at all times, i.e. $y_{j+1}\geq y_j+\Delta$, for $j=1,\ldots
,N-1$.  The end of the system are reflecting (the particles cannot
escape), i.e.  $y_1>-(L-\Delta)/2$ and $y_N<(L-\Delta)/2$.
 
The diffusion of finite-sized particles can be mapped onto a
point-particle problem. Introducing the rescaled effective system
length
\be\label{eq:PPmap1}
        \ell=L-N\Delta,
\ee
and making the coordinate transformation 
\bea\label{eq:PPmap2}
        x_j&=&y_j-j\Delta + \frac{N+1}{2}\Delta,\nonumber\\
        x_{j,0}&=&y_{j,0}-j\Delta + \frac{N+1}{2}\Delta,
\eea
leads to 
\be\label{eq:region_R}
         {\cal R}:Ê-\ell/2<x_1<x_2\ldots <x_N<\ell/2 
\ee
where ${\cal R}$ denotes the phase space spanned by
(\ref{eq:region_R}).  The phase space ${\cal R}_0$ is also introduced
for the initial coordinates which satisfy
$-\ell/2<x_{1,0}<x_{2,0}\ldots <x_{N,0}<\ell/2$. For convenience, we
also introduce the short-hand notation $\vec{x}= (x_1,\ldots,x_N)$ and
$\vec{x}_0 = (x_{1,0},\ldots,x_{N,0})$. Equations (\ref{eq:PPmap1})
and (\ref{eq:PPmap2}) maps {\em exactly} the problem of $N$
finite-sized hard-core particles in a box of length $L$ onto a $N$
point-particle problem in a box of length $\ell$.

The main quantity of interest in this study is the tPDF $\rho_{\cal
  T}(x_{\cal T},t|x_{{\cal T},0})$, that is the probability density
that a tagged particle ${\cal T}$ (${\cal T}=1,...,N$) is at position
$x_{\cal T}$ at time $t$, given that it was at $x_{{\cal T},0}$ at
$t=0$ [an ensemble average over over the initial (equilibrium)
distribution of the surrounding $N-1$ particles is implicit]. The
equilibrium tPDF is straightforwardly calculated from the ergodicity
principle: all points in the allowed phase space ${\cal R}$ are
equally probable. This leads the equilibrium $N$PDF
\be\label{eq:P_N_eq} 
        {\cal P}^{\rm eq}(\vec{x})
        =\frac{N!}{\ell^N} \prod_{i=1}^{N-1} \theta (x_{i+1}-x_i), 
\ee 
where $\theta(z)$ is the Heaviside step function, $\theta(z)=1$ for
$z>0$ and zero elsewhere. Using an extended phase space integration
technique (see Appendix \ref{sec:extended_int}) it is easy to verify
that
$\int_{\cal R} dx_1 \cdots dx_N {\cal P}^{\rm eq}(\vec{x})=1$.
Integrating Eq. (\ref{eq:P_N_eq}) over all coordinates leaving out
one, $x_{\cal T}$, gives the tPDF
\bea\label{eq:P_eq_first}
         \rho_{\cal T}^{\rm eq}(x_{\cal T}) &=& \int_{\cal R} dx_1'\cdots dx_N'         \delta(x_{\cal T} -x_{\cal T}') {\cal P}^{\rm eq}(\vec{x}')\nonumber\\
&=&\frac{N!}{\ell^N} \frac{1}{N_L!\,N_R!} \int_{-\ell/2}^{x_{\cal T}} dx_1'
\cdots \int_{-\ell/2}^{x_{\cal T}} dx_{{\cal T}-1}' \nonumber\\
&&\times \int_{x_{\cal T}}^{\ell/2} dx_{{\cal T}+1}' \cdots \int_{x_{\cal T}}^{\ell/2} dx_N' \nonumber\\
&=&\frac{1}{\ell^N} \frac{N!}{N_L!\,N_R!}
  \left(\frac{\ell}{2}+x_{\cal T}\right)^{N_L}  \left(\frac{\ell}{2}-x_{\cal T}\right)^{N_R},
  \eea
where $\delta(z)$ is the Dirac delta-function and $N_L$ ($N_R$) is
the number of particles to the left (right) of the tagged particle
($N=N_L+N_R+1$).  In the remaining part of this section, we show how
to calculate the complete time evolution of the tPDF from the
many-particle $N$PDF.

In order to obtain $\rho_{\cal T}(x_{\cal T},t|x_{{\cal T},0})$ one
needs to first introduce the $N$-particle {\em joint} probability
density ${\cal P}(\vec{x},t;\vec{x}_0)$ which gives the probability
density that the system is in a state $\vec{x}$ {\em and} that it
initially was in a state $\vec{x}_0$. The joint probability density
for the tagged particle $\rho_{\cal T}(x_{\cal T},t;x_{{\cal T},0})$ is simply
obtained from the joint $N$PDF by integration over ${\cal R}$ and
${\cal R}_0$:
$\rho_{\cal T}(x_{\cal T},t;x_{{\cal T},0})=
\int_{\cal R} dx_1'\cdots dx_N' \int_{{\cal R}0} dx_{1,0}' 
\cdots dx_{N,0}'\delta(x_{\cal T} -x_{\cal T}') \delta(x_{{\cal
T},0} -x_{{\cal T},0}') {\cal P}(\vec{x}',t;\vec{x}_0')$.
Relating the conditional and joint probability densities using Bayes'
rule \cite{VanKampen}, i.e.
${\cal P}(\vec{x},t;\vec{x}_0)=
{\cal P}(\vec{x},t|\vec{x}_0){\cal P}^{\rm eq}(\vec{x}_0)$
and 
$\rho_{\cal T}(x_{\cal T},t;x_{{\cal T},0})=
\rho_{\cal T}(x_{\cal T},t | x_{{\cal T},0})\rho^{\rm eq}(x_{{\cal T},0})$, 
leads to
\bea\label{eq:OnepartDistr}
        \rho_{\cal T}(x_{\cal T},t|x_{{\cal T},0})&=&
        \frac{1}{\rho^{\rm eq}(x_{{\cal T},0})} 
        \int_{\cal R} dx_1'\cdots dx_N' \delta(x_{\cal T} -x_{\cal T}')
        \nonumber\\
        &&\int_{{\cal R}0} dx_{1,0}' \cdots  dx_{N,0}'\delta(x_{{\cal
        T},0} - x_{{\cal T},0}') \nonumber\\
        & & \times   {\cal P}(\vec{x}',t|\vec{x}_0') P^{\rm eq}(\vec{x}_0'),
\eea
where $\rho^{\rm eq}_{\cal T}(x_{{\cal T},0})$ is given in Eq.
(\ref{eq:P_eq_first}).

In order to get the tPDF using Eq.
(\ref{eq:OnepartDistr}), we need to calculate the $N$PDF ${\cal
  P}(\vec{x},t|\vec{x}_0)$. It  is governed by the diffusion equation
\be\label{eq:DiffEq}
 \frac{ \partial {\cal P}(\vec{x},t|\vec{x}_0)}{\partial t} = 
   D \left(
    \frac{\partial^2}{\partial x_1^2}+\frac{\partial^2}{\partial x_2^2}+ \ldots +\frac{\partial^2}{\partial x_N^2}
  \right)
  {\cal P}(\vec{x},t|\vec{x}_0),
\ee
for $\vec{x}\in {\cal R}$ (${\cal P}(\vec{x},t|\vec{x}_0)\equiv 0$
outside ${\cal R}$). The equation certifying that neighboring particles
cannot overtake reads
\be\label{eq:NopassCond}
  \left. D \left(
      \frac{\partial}{\partial x_{i+1}}- \frac{\partial}{\partial x_i}
   \right)
   {\cal P}(\vec{x},t|\vec{x}_0)\right|_{x_{i+1}=x_i} = 0.
\ee
Also, reflecting boundaries are placed at the system  ends 
\be\label{eq:NofluxCond1}
 D \left. \frac{\partial{\cal P}(\vec{x},t|\vec{x}_0)}{\partial x_1}\right|_{x_1=-\ell/2}= 0,
 \ee
\be\label{eq:NofluxCond2}
 D \left. \frac{\partial{\cal P}(\vec{x},t|\vec{x}_0)}{\partial x_N}\right|_{x_N=\ell/2}=0,
\ee
 making sure that the particles are restricted to $[-\ell/2,\ell/2]$ at all times. Finally, the initial condition is
\be\label{eq:InitCond}
 {\cal P}(\vec{x},0|\vec{x}_0) = \delta(x_1-x_{1,0})\cdots \delta(x_N-x_{N,0}).
\ee

Summarizing this section, the problem of $N$ hard-core interacting
particles of size $\Delta$ diffusing in a one-dimensional system of a
finite length $L$ was mapped onto a point-particle problem using
relationships (\ref{eq:PPmap1}) and (\ref{eq:PPmap2}). The dynamics of
the $N$PDF is governed by
Eqs.~(\ref{eq:DiffEq})-({\ref{eq:NofluxCond2}}). Once they are solved
(topic of Sec. \ref{sec:NpartDist}), the tPDF can be calculated via
Eq. (\ref{eq:OnepartDistr}) (as demonstrated in Sec.
\ref{sec:P_bar_general}).


\section{$N$PDF as a coordinate Bethe ansatz}\label{sec:NpartDist}

In this section we obtain the $N$PDF for the diffusion problem defined
in previous section using a coordinate Bethe ansatz. The Bethe ansatz
has been proven useful in solving a large variety of interacting
particle problems since its introduction by Hans Bethe in 1931 (see
\cite{Batchelor_07} for a review). The Bethe ansatz solution for the
present problem reads
\bea\label{eq:BetheAnsatz}
  {\cal P}(\vec{x},t|\vec{x}_0) &=&
  \int_{-\infty}^\infty  \frac{dk_1}{2\pi}  \int_{-\infty}^\infty  
\frac{dk_2}{2\pi}
  \cdots \int_{-\infty}^\infty \frac{dk_N}{2\pi}\nonumber\\
& \hspace{-1cm} \times &  \hspace{-0.5cm}
        e^{-E(k_1,\ldots,k_N) t}\Pi_{j=1}^N \phi (k_j,x_{j,0}) \nonumber\\
  & \hspace{-1cm} \times & \hspace{-0.5cm}
          \big[ e^{ik_1x_1+ik_2x_2+ik_3x_3+\ldots+ik_Nx_N} \nonumber \\
  & \hspace{-0.8cm} + & \hspace{-0.5cm}
          S_{21}e^{ik_2x_1+ik_1x_2+ik_3x_3+\ldots+ik_Nx_N} \nonumber \\
  & \hspace{-0.8cm} + & \hspace{-0.5cm}
          S_{32}S_{31}e^{ik_2x_1+ik_3x_2+k_1x_3+ \ldots + ik_Nx_N}  \nonumber \\
  & \hspace{-0.8cm} + &\hspace{-0.5cm}  { \rm all \ other \ permut. \ of  
\ } \{ k_1,k_2,...,k_N\}\big]
\eea
where $E(k_1,\ldots,k_N)$ is the dispersion relation, $S_{ij}$ are the
scattering coefficients, and $\phi(k_j,x_{j,0})$ denotes a function
containing boundary and initial conditions. Each one of these
quantities are described below.

The dispersion relation has the form
\be\label{eq:disp_rel}
  E(k_1,\ldots,k_n) = D(k_1^2+\ldots+k_N^2),
\ee
and relates "energy" to the momenta $k_1,\ldots,k_N$. Equation  
(\ref{eq:disp_rel}) is obtained by inserting the Bethe ansatz Eq.  
(\ref{eq:BetheAnsatz}) into the equation of motion, Eq.  
(\ref{eq:DiffEq}).

The scattering coefficients $S_{ij}$ describe pair-wise particle  
interactions and are in general functions of the momentum variables  
$k_i$ and $k_j$, $S_{ij} = S(k_i,k_j)$. They  are, however,   
independent of the initial positions of the particles.
In Appendix \ref{sec:Bethe_ansatz_proofs} it is demonstrated that the  
scattering coefficients making sure that the particles cannot pass each other  
[i.e. satisfying Eq. (\ref{eq:NopassCond})]  are given by
\be\label{eq:S_ij}
        S_{ij}=1,
\ee
which means that they are independent of momenta and correspond  
to perfect reflection. For non-interacting particles $S_{ij}=0$, which  
reduces the Bethe ansatz to a standard Fourier transform. 

The quantity $\phi(k_j,x_{j,0})$ contains information about the initial  
and boundary conditions of the problem, here defined by Eqs.  
  (\ref{eq:NofluxCond1}) - (\ref{eq:InitCond}). 
The form of $\phi(k_j,x_{j,0})$ satisfying  
these relationships is given by
\be\label{eq:I0_refl}
  \phi (k_j,x_{j,0}) = 2 \cos [k_j(x_{j,0}+\ell/2)] \sum_{m=-\infty}^\infty
      e^{ik_j(2m+1/2)\ell},
\ee
which is shown explicitly in Appendix \ref{sec:Bethe_ansatz_proofs}.  
Notably,  for an infinite system we have $\phi (k_j,x_{j,0})=e^{- 
ik_jx_{j,0}}$ \cite{footnote2,footnote3}.

It is interesting to note that for SFD systems described by Eq.
(\ref{eq:BetheAnsatz})-(\ref{eq:I0_refl}) any {\em macroscopic}
quantity that is invariant under interchange of any two particle
positions, $x_i\leftrightarrow x_j$, takes the same value as for a
system of non-interacting particles. This is in marked contrast to
{\em microscopic} quantities such as the tPDF which in general behave
very differently for single-file and independent particle systems. In
Appendix \ref{sec:macroscopic} we use the Bethe ansatz to explicitly
calculate two macroscopic quantities, the dynamic structure factor and
the center of mass PDF, and show that they agree with standard results
for independent particle systems.

Integration over momenta in Eq. (\ref{eq:BetheAnsatz}) [using Eqs. (\ref{eq:disp_rel})-(\ref{eq:I0_refl})] leads to the  $N$PDF:
\begin{eqnarray}\label{eq:BetheAnsatz2}
  {\cal P}(\vec{x},t|\vec{x}_0) &=&
  \psi(x_1,x_{1,0};t)\psi(x_2,x_{2,0};t)\cdots\psi(x_N,x_{N,0};t)\nonumber 
\\
  &\hspace{-1cm}+& \hspace{-0.5cm} \psi(x_1,x_{2,0};t)\psi(x_2,x_{1,0};t) 
\cdots\psi(x_N,x_{N,0};t)\nonumber \\
  &\hspace{-1cm}+& \hspace{-0.5cm} { \rm all \ other \ permut. \ of  
\ }  \{x_{1,0},x_{2,0},...,x_{N,0} \},
\end{eqnarray}
where
\bea\label{eq:P_single_refl}
   \psi(x_i,x_{j,0};t) &=&
  \frac{1}{(4\pi D t)^{1/2}} \nonumber \\
  &\hspace{-1cm} \times& \hspace{-0.5cm}\sum_{m=-\infty}^\infty
  \left\{ \exp\left[-\frac{(x_i-x_{j,0}+2m\ell)^2}{4Dt}\right] \right.
  \nonumber \\
&\hspace{-1cm}+& \hspace{-0.5cm}\left. \exp\left[-\frac{(x_i+x_{j,0}+(2m 
+1)\ell)^2}{4Dt}\right] \right\},
\eea
is obtained from the inverse Fourier transform
$
(2\pi)^{-1}\int_{-\infty}^{\infty} dk_j\, \phi(k_j,x_{j,0})
e^{- Dk_j^2t}e^{ik_j x_i}.
$
We point out that  $\psi(x_i,x_{j,0};t)$ is the single-particle PDF
for a particle in confined in a box of length $\ell$.

The single-particle PDF given in Eq.  (\ref{eq:P_single_refl}) is,
however, not convenient for analyzing the long time limit
$t\rightarrow \infty$.  In order to get a more suitable expression we
seek instead the eigenmode expansion of $ \psi(x_i,x_{j,0};t)$, which
can be done in a variety of ways. Here we use Bromwich integration.
The Laplace transform of Eq. (\ref{eq:P_single_refl}) is (see
Appendix \ref{sec:eigenmode_exp})
\bea\label{eq:Psi_s2}
        \psi(x_i,x_{j,0};s)&=&\int_0^\infty dt \, e^{-st} \psi(x_i,x_{j,0};t) \nonumber\\
&=&\frac{1}{\sqrt{4Ds} \sinh(\ell \sqrt{s/D})}   
\nonumber\\
        &\times& \left(
        \cosh [(x_i+x_{j,0})\sqrt{s/D}] \right.
        \nonumber \\
    &&+ \left. \cosh [(\ell-|x_{j,0}-x_i|)\sqrt{s/D}]\right).
   \eea
The sought eigenvalue expansion is obtained as a sum of residues of $\psi(x_i,x_{j,0};s)$ (see e.g. Ref. \cite{Wunsch}), and reads
\be\label{eq:P_single_refl2}
   \psi(x_i,x_{j,0};t)=\frac{1}{\ell}\Big[ 1+ 
   \sum_{m=1}^\infty  {\cal G}_m(x_i,x_{j,0})\,{\cal E}_m(t)\Big]
\ee
where
\bea\label{eq:G_p}
    {\cal G}_m(x_i,x_{j,0}) &=&
     \nu^{(+)}_m \cos \left(\frac{m\pi x_i}{\ell}\right) 
     \cos \left(\frac{m\pi x_{j,0}}{\ell}\right) \nonumber\\
     &&+\nu_m^{(-)} \sin \left( \frac{m\pi x_i}{\ell}\right) 
     \sin \left(\frac{m\pi x_{j,0}}{\ell}\right),
\eea
\be\label{eq:F_p}
{\cal E}_m(t)=e^{- (m\pi)^2 D t /\ell^2}
   \ee
and
\bea\label{eq:cp_sp}
        \nu_m^{(\pm)}&=&1\pm(-1)^m.
\eea
Elementary trigonometric identities \cite{ABST} were used to bring
${\cal G}_m(x_i,x_{j,0})$ onto the form in Eq. (\ref{eq:G_p}).
Equations (\ref{eq:P_single_refl2})-(\ref{eq:cp_sp}) agrees with
well-known results \cite{Carslaw}. The single-particle PDF
(\ref{eq:P_single_refl2}) is more convenient for obtaining the long
time limit as well as for numerical computations compared to Eq.
(\ref{eq:P_single_refl}).

In summary, the many-particle $N$PDF for excluding particles of size $
\Delta$ diffusing in a finite interval of length $L$ with reflecting
boundaries is given by Eqs.~(\ref{eq:BetheAnsatz2}),
(\ref{eq:P_single_refl2}) [or Eq. (\ref{eq:P_single_refl})] combined
with the mapping equations (\ref{eq:PPmap1}) and (\ref{eq:PPmap2}).
For point-particles ($\Delta=0$) these results agree with those
presented in \cite{RKH} where a different approach was
used~\cite{footnote1}. Based on the explicit expression of our $N$PDF,
we will in the following section address the tPDF.


\section{\lowercase{t}PDF - exact and large $N$ results}\label{sec:P_bar_general}

In this section, we calculate the tPDF (\ref{eq:OnepartDistr}) by integrating out the coordinates and initial positions of all non-tagged particles from the $N$PDF given in  Eq. (\ref{eq:BetheAnsatz2}) \cite{Aslangul_connection}.
As is shown in detail in Appendix \ref{sec:extended_int}
we can, due to the property that ${\cal P}(\vec{x},t|\vec{x}_0)$
is invariant under permutations of $x_i\leftrightarrow x_j$, extend the
integration from ${\cal R}$ to the hypercubes $x_j\in[-\ell/2,x_{\cal T}],\ j=1,...,{\cal T}-1$ and $x_j\in [x_{\cal T},\ell/2], \
j={\cal T}+1,...,N$. A similar procedure holds for integration over ${\cal R}_0$ (initial positions). Using the extended phase-space technique, Eq. (\ref{eq:OnepartDistr})
becomes
\bea\label{eq:P_bar}
&&\rho_{\cal T}(x_{\cal T},t|x_{{\cal T},0})=\frac{f_L^{N_L}f_R^{N_R}}{N_L!N_R!}  \int_{-\ell/2}^{x_{\cal T}} dx_1\cdots
\int_{-\ell/2}^{x_{\cal T}} dx_{{\cal T}-1} \nonumber \\
&&
\times \int_{x_{\cal T}}^{\ell/2} dx_{{\cal T}+1} 
\cdots \int_{x_{\cal T}}^{\ell/2} dx_N
\int_{-\ell/2}^{x_{{\cal T},0}} dx_{1,0}  
\cdots\\
&&\times
  \int_{-\ell/2}^{x_{{\cal T},0}} dx_{{\cal T}-1,0}
  \int_{x_{{\cal T},0}}^{\ell/2} dx_{{\cal T}+1,0} \cdots 
  \int_{x_{{\cal T},0}}^{\ell/2} dx_{N,0} {\cal P}(\vec{x},t|\vec{x}_0),        
  \nonumber
\eea
where 
\be \label{eq:fL_fR}
f_L=(\ell/2+x_{{\cal T},0})^{-1},\ \ \
f_R=(\ell/2-x_{{\cal T},0})^{-1}.
\ee
Using a similar combinatorial analysis to the one in Ref. \cite{RKH}, we arrive at  Eq. (\ref{eq:P_bar2}) (see  Appendix \ref{sec:PDF_integration}). The tPDF is, however, more conveniently expressed in terms of Jacobi polynomials \cite{ABST}, $P_n^{(\alpha,\beta)}(z)$. Using the identities 
$P_n^{(\alpha,\beta)}(z)=(n+\alpha)!(n+\beta)!/[n!(n+\alpha+\beta)!]
[(z-1)/2]^{-a} P_{n+\alpha}^{(-\alpha,\beta)}(z)$ and
$P_n^{(\alpha,\beta)}(-z)=(-1)^nP_n^{(\beta,\alpha)}(z)$ \cite{Szego_59} leads to
\bea\label{eq:P_bar3}
&&\rho(x_{\cal T},t|x_{{\cal T},0})=\frac{(N_L+N_R-1)!}{N_L!N_R!} (\psi_L^L)^{N_L} (\psi_R^R)^{N_R} \nonumber\\
&& \times \Big\{ (N_L+N_R) \psi \Phi(0,0,0;\xi)
+ N_L^2 \frac{\psi^L\psi_L}{\psi_L^L} \Phi(1,0,0;\xi)\nonumber\\
&&\ \ +N_R^2 \frac{\psi^R\psi_R}{\psi_R^R} \Phi(0,1,0;\xi)\nonumber\\
&&\ \ +N_L N_R \left[
        \frac{\psi^R\psi_L}{\psi_L^R}+ \frac{\psi^L\psi_R}{\psi_R^L}
         \right] \Phi(0,0,1;\xi) \Big\}
\eea
where
\bea\label{eq:Phi_main}
\Phi(a,b,c;\xi)&=&\frac{(N_L-(a+c))!(N_R-b)!}{(N_L+N_R-(a+b+c))!} \nonumber\\
        & &\hspace{-2cm} \times \,  
        \xi^c (1-\xi)^{N_L-(a+c)} P_{N_L-(a+c)}^{(c,N_R-N_L+a-b)}\left(\frac{1+\xi}{1-\xi}\right)
\eea
and  
\be\label{eq:xi}
        \xi=\frac{\psi_R^L \psi_L^R}{\psi_L^L\psi_R^R}.
\ee
were introduced. The quantities $\psi_L$, $\psi_L^L$ etc. are defined in
Eq. (\ref{eq:integrals_def}) and are integrals of the one
particle propagator $\psi=\psi(x_i,x_{j,0};t)$ \cite{rel_to_rodenbeck_pdf}.
For the general case we have $\xi \in [0, 1]$ \cite{xi_y}. By using
limiting results of $\psi_L^L$, $\psi_R^R$, $\psi_L^R$ and $\psi_R^L$
from Appendix \ref{sec:Laplace_integrals}, one concludes that
$\xi\rightarrow 0$ for short times ($t\rightarrow 0$), and $\xi
\rightarrow 1$ in the long time limit ($t\rightarrow \infty$). It is
also possible using standard relations for the Jacobi polynomials
\cite{ABST} to express $\Phi(a,b,c;\xi)$ as a Gauss hypergeometric
function ${}_2F_1(\alpha,\beta,\gamma;z)$:
\bea\label{eq:Phi2_main}
        \Phi(a,b,c;\xi)&=&
{ }_2F_1(-N_L+a,-N_R+b, \nonumber \\ &&-(N_L+N_R)+a+b+c;1-\xi)
\eea
which is convenient for obtaining the long-time behavior as will be
seen in the next section.

Finally, we need to find explicit expressions for the integrals of
$\psi(x_i,x_{j,0};t)$ [Eq. (\ref{eq:integrals_def})].
Integrating Eq. (\ref{eq:P_single_refl2}) [or integrating
$\psi(x_i,x_{j,0};s)$ prior to Laplace inversion, see Appendix
\ref{sec:Laplace_integrals}] yields (arguments left implicit)
  \bea\label{eq:Psi_LL_etc}
\psi_L^L&=&\frac{1}{2}+\frac{x_{\cal T}}{\ell}+\left(\frac{1}{2}+\frac{x_{{\cal T},0}}{\ell}\right)^{-1} \sum_{m=1}^\infty  K_m \,{\cal E}_m(t),\nonumber\\
\psi_R^R&=&\frac{1}{2}-\frac{x_{\cal T}}{\ell}+\left(\frac{1}{2}-\frac{x_{{\cal T},0}}{\ell}\right)^{-1} \sum_{m=1}^\infty K_m \,{\cal E}_m(t),\nonumber\\
\psi^L&=&\frac{1}{2}+\frac{x_{\cal T}}{\ell}+ \sum_{m=1}^\infty J_m (x_{\cal T},x_{{\cal T},0}) {\cal E}_m(t),\nonumber\\
\psi_L&=&\frac{1}{\ell}\left[ 1 + \left(\frac{1}{2}+\frac{x_{{\cal T},0}}{\ell}\right)^{-1} \sum_{m=1}^\infty J_m(x_{{\cal T},0},x_{\cal T}) \,{\cal E}_m(t)\right],\nonumber\\
\psi_R&=&\frac{1}{\ell} \left[1 - \left(\frac{1}{2}-\frac{x_{{\cal T},0}}{l}\right)^{-1} \sum_{m=1}^\infty J_m(x_{{\cal T},0},x_{\cal T}) \,{\cal E}_m(t)\right],\nonumber\\
\psi^R&=&1-\psi^L , \ \psi_L^R=1-\psi_L^L, \ \psi_R^L=1-\psi_R^R
  \eea
with
  \bea\label{eq:J_K}
K_m&=&\frac{1}{(m\pi)^2} \left[ 
                \nu_m^{(+)} \sin \left( \frac{m\pi x_{\cal T}}{\ell} \right) 
                \sin \left(\frac{m\pi x_{{\cal T},0}}{\ell}\right) \right. \nonumber\\
&& \left.+ \nu_m^{(-)} \cos \left(\frac{m\pi x_{\cal T}}{\ell}\right) 
         \cos \left(\frac{m\pi x_{{\cal T},0}}{\ell}\right) \right] \\
J_m(z,z')&=&\frac{1}{m\pi} \left[
        \nu_m^{(+)} \sin \left(\frac{m\pi z}{\ell} \right) 
        \cos \left(\frac{m\pi z'}{\ell}\right) \right.\nonumber\\
&& \left. - \nu_m^{(-)} \cos \left(\frac{m\pi z}{\ell}\right) 
         \sin \left(\frac{m\pi z'}{\ell}\right)\right],
  \eea
where ${\cal E}_m(t)$ and $\nu_m^{(\pm)}$ are given by Eqs.
(\ref{eq:F_p}) and (\ref{eq:cp_sp}), respectively.
To summarize, the complete expression for the tagged particle PDF is given by
Eqs. (\ref{eq:P_bar3})-(\ref{eq:xi}) and (\ref{eq:Psi_LL_etc})-(\ref{eq:J_K}) together with Eqs. (\ref{eq:PPmap1}) and (\ref{eq:PPmap2}) for the case of
finite-sized particles. The expressions for $\rho_{\cal T}(x_{\cal T},t|x_{{\cal
T},0})$ can straightforwardly be computed numerically; our
\texttt{MATLAB} implementation is available upon~request.

In the remaining part of this section we derive the tagged particle
PDF $\rho_{\cal T}(x_{\cal T},t|x_{{\cal T},0})$ valid for a large $N$
and (finite) system size $\ell$. This large $N$ expansion will be used
in the next section for identifying different time regimes and to
obtain $\rho_{\cal T}(x_{\cal T},t|x_{{\cal T},0})$ for short and
intermediate times.  From Eq. (\ref{eq:Phi_main}) we note that the
argument in the Jacobi polynomial, $P_n^{(\alpha,\beta)}(z)$, is in
the interval $z\in [1, \infty)$ (since $\xi \in [0,1]$) and that the
number of particles is related to the order $n$. A large $N$ expansion
of $\Phi(a,b,c;\xi)$ therefore amounts to find a large order $n$
expansion valid for $z\in [1,\infty)$ (i.e. for all times) for the
Jacobi polynomial. One such expansion was derived in \cite{Elliott_71}
(see also \cite{Olver_54,Olver_56}), and applying it to Eq.
(\ref{eq:Phi_main}) yields
  \bea\label{eq:Phi_Bessel}
\Phi(a,b,c;\xi)&\approx &[N-(a+b+c)]^{1/2}\xi^{(2c-1)/4}\nonumber\\
&&\times (2\pi \zeta)^{1/2}\left(\frac{1-\xi}{4}\right)^{[N-(a+b+c)]/2}\nonumber\\
&&\times  I_c[(N-(a+b+c))\zeta]\big(1+A(\zeta)\big)\nonumber\\
  \eea
where \cite{zeta_def}
 \be\label{eq:zeta}
\zeta=\frac{1}{2} \ln \left[ \frac{1+\sqrt{\xi}}{1-\sqrt{\xi}}\right],
  \ee
  and $I_\alpha(z)$ is the modified Bessel function of the first kind
  of order $\alpha$. Stirling's formula \cite{ABST} was also used to
  approximate factorials involving $N_L$ and $N_R$.  The correction
  term appearing in Eq. (\ref{eq:Phi_Bessel}) is
  \bea
A(\zeta)&=&\frac{B_0(\zeta)}{N-(a+b+c)}\frac{I_{c+1}[(N-(a+b+c))\zeta]}{I_c [(N-(a+b+c))\zeta]}\nonumber\\
B_0(\zeta)&=&\frac{1}{2} \Big\{ \left(c^2-1/4\right)
\left(\frac{1}{\zeta}-\coth{\zeta}\right)\nonumber\\
&& -[(\delta_N+a-b)^2-1/4]\tanh(\zeta) \Big \},
  \eea
where $\delta_N=N_R-N_L $ was introduced. The expression for
$A(\zeta)$ was found by explicitly evaluating an integral in
Ref. \cite{Elliott_71}, and it is a straightforward matter to show that
the correction term $A(\zeta)$ is indeed always small  for all $\xi\in
[0, 1]$ provided that $1/N,|\delta_N|/N\ll 1$. Inserting
Eq. (\ref{eq:Phi_Bessel}) in Eq. (\ref{eq:P_bar3}) and using the same
approximations as above, we obtain our final large $N$ result for the
tPDF:
  \bea\label{eq:P_bar_Bessel}
&&\rho_{\cal T}(x_{\cal T},t|x_{{\cal T},0})=(\psi_L^L)^{(N-\delta_N-1)/2} (\psi_R^R)^{(N+\delta_N-1)/2}\nonumber\\
&& \ \ \times (1-\xi)^{(N-1)/2} \left( \frac{\zeta}{\sqrt{\xi}}\right)^{1/2}
%
\Big\{ (1-\xi)^{1/2} \psi I_0[N\zeta]
 \nonumber\\ 
&&
\ \ + \ \frac{N}{2} \left(
        \frac{\psi^L\psi_L}{\psi_L^L}+\frac{\psi^R\psi_R}{\psi_R^R} 
         \right) I_0[N\zeta]
\nonumber\\
&& 
 \ \ + \ \frac{N}{2} \sqrt{\xi} \left(
        \frac{\psi^R\psi_L}{\psi_L^R}+\frac{\psi^L\psi_R}{\psi_R^L}
        \right) I_1[N\zeta] \Big\}.
  \eea
We point out that this expression, in contrast to previous asymptotic
expressions \cite{LE,Hahn_95,RKH}, is valid for a {\em finite} box of size
$\ell$, assuming only that the number of particles is large $N\gg 1$ and that
the tagged particle is approximately in the center of the system:
$|\delta_N|/N\ll 1$.


\section{Three different time regimes}
\label{sec:three_regimes}

In this section we show that for large $N$, the finite SFD system
considered here has three different time regimes to which expressions
for $\rho_{\cal T}(x_{\cal T},t|x_{{\cal T},0})$ are derived.
Mathematically, the different cases appear due to the magnitude of
$N\zeta$ [found in the argument of the Bessel functions in Eq.
(\ref{eq:P_bar_Bessel})], and if $\zeta$ is small or large. Utilizing
Eqs. (\ref{eq:Psi_LL_etc_short_time}) and (\ref{eq:zeta}), these cases
can be turned into different time regimes if introducing the
collision time
\be\label{eq:t_coll}
        \tau_{\rm coll}=\frac{1}{\varrho^2D},
\ee
where $\varrho=N/\ell$ is the concentration of particles, and the equilibrium time 
\be\label{eq:t_eq}
  \tau_{\rm eq}=\frac{\ell^2}{D}.
\ee
For a particle located roughly in the middle, $|\delta_N|/N\ll 1$ (i.e. 
Eq. (\ref{eq:P_bar_Bessel}) applies), the three cases are  given by
\begin{enumerate}
  \item[A.] {\em short} times, $N\zeta\ll 1$, 
                        i.e. $t\ll \tau_{\rm coll},\tau_{\rm eq}$,
  \item[B.] {\em intermediate} times, $\zeta\ll1$ and $N\zeta\gg 1$,    
                        corresponding to $\tau_{\rm coll}\ll t \ll \tau_{\rm eq}$, 
  \item[C.] {\em long} times, $\zeta\gg 1$, i.e. $t\gg \tau_{\rm coll},
                        \tau_{\rm eq}$.

\end{enumerate}
Time regimes (A)-(C) are analyzed in detail below.

\subsection{Short times, $t\ll \tau_{\rm coll},\tau_{\rm eq} $}
For short times we have $N\zeta\ll 1$, and may therefore use the
approximations $I_\alpha(z)|_{z \ll 1} \approx
(z/2)^\alpha/\Gamma(\alpha+1)$ \cite{ABST}, $\zeta\approx \sqrt{\xi}$
and $\xi\ll 1$. In this limit, one finds that the first term in Eq.
(\ref{eq:P_bar_Bessel}) dominates which in combination with Eq.
(\ref{eq:PPmap2}) leads to
\be\label{eq:P_bar_t0}
        \rho_{\cal T}(y_{\cal T},t|y_{{\cal T},0})=(4\pi D t)^{-1/2} 
        \exp\left[-\frac{(y_{\cal T}-y_{{\cal T},0})^2}{4Dt}\right],
\ee
for which the MSD is
\be
        {\cal S}(t) =2Dt.
\ee
In the short time regime, almost no collisions with the neighboring particles
(nor the box walls) have occurred and the tPDF is
therefore a Gaussian with width $2Dt$ as for a free particle in an
infinite one-dimensional system.

\subsection{Intermediate times, $\tau_{\rm coll} \ll t \ll \tau_{\rm eq}$}\label{sec:intermediate}

In the intermediate time regime the tagged particle has collided many times with its neighbors but not yet reached its equilibrium tPDF. For this regime, where $\zeta\ll 1$ but $N\zeta\gg 1$, we get the tPDF as follows. First,   the first term in Eq. (\ref{eq:P_bar_Bessel}) is neglected (this is checked at the end of the calculation). Second, the Bessel function is approximated with
$I_\alpha(z)|_{z\gg 1}\approx e^z/\sqrt{2\pi z}$. A
straightforward expansion of Eq. (\ref{eq:P_bar_Bessel}) for
$\psi_L^R,\psi_R^L\ll 1$ (i.e. $\sqrt{\xi}\ll 1$), together with
Stirling's formula, gives:
%
%
\bea\label{eq:P_bar_Psi_LR_small}
&&\rho_{\cal T}(y_{\cal T},t|y_{{\cal T},0})\approx \frac{1}{2} 
        e^{-\delta_N \left(\psi_R^L-\psi_L^R \right)}
        \sqrt{\frac{N}{2\pi}}
        \left(\psi_R^L\psi_L^R\right)^{-1/4}\nonumber\\
&&\ \times\, e^{-\frac{N}{2}\left(\sqrt{\psi_R^L}-\sqrt{\psi_L^R}\right)^2}  
        \Big[\psi^L\psi_L+\psi^R\psi_R
        \nonumber\\
&& \ \  \ +\, \psi^R \psi_L \left(\frac{\psi_R^L}{\psi_L^R}\right)^{1/2}
        +\psi^L \psi_R \left(\frac{\psi_L^R}{\psi_R^L}\right)^{1/2}
        \Big].
  \eea
If we furthermore assume that the average of the absolute value of
$\eta=(x_{\cal T}-x_{{\cal T},0})/\sqrt{4Dt}$ is small (which is checked
after the calculation), Eq. (\ref{eq:Psi_LL_etc_eta_expansion}) may be used which in combination with Eqs. (\ref{eq:P_bar_Psi_LR_small}), (\ref{eq:PPmap1}) and (\ref{eq:PPmap2}),  keeping only lowest order terms in $\eta$, leads to the SFD result:
\bea\label{eq:P_bar_SFD}
        &&\rho_{\cal T}(y_{\cal T},t|y_{{\cal T},0})= 
        \frac{1}{\sqrt{2\pi}}   \left(\frac{1}{\frac{4Dt}{\pi} (\frac{1-        \varrho \Delta}{\varrho})^2} \right)^{1/4} \nonumber\\
&&\ \ \ \times
 \exp \left[-\frac{(y_{\cal T}-y_{{\cal T},0})^2}{2 \sqrt{ \frac{4Dt}{\pi}      (\frac{1-\varrho\Delta}{\varrho})^2 } }\right]
\eea
where the MSD is
\be\label{eq:MSD_SFD}
        {\cal S}(t)=\frac{1-\varrho\Delta}{\varrho}\sqrt{\frac{4Dt}{\pi}}.
\ee
Equation (\ref{eq:MSD_SFD}) justifies the assumption that the
expectation value of $|\eta|$ is a small number. Also, comparing the
magnitude of the first term in Eq. (\ref{eq:P_bar_Bessel}) with respect
to the second and the third, shows indeed that our first assumption
above was correct \cite{some_details}. For point-particles $\Delta=0$,
Eq. (\ref{eq:P_bar_SFD}) agrees with standard results \cite{LE,RKH},
in which $N,L\rightarrow\infty$ while keeping the concentration
$\varrho$ fixed. The result above shows that SFD behavior appears also
in a finite system with reflecting ends, as an intermediate regime
(for a particle roughly in the middle). In addition, note that the
simple rescaling $\varrho\rightarrow \varrho/(1-\varrho \Delta)$ takes
us from previous point-particle results to those of finite-sized
particles.

\subsection{Long times, $t\gg \tau_{\rm eq}$}

In the long-time limit we have $\zeta\gg 1$, for which the tPDF can be obtained exactly for arbitrary $N$. Using the exact expression for $\rho_{\cal
T}(x_{\cal T},t|x_{{\cal T},0})$ found in Eqs. (\ref{eq:P_bar3}) and
(\ref{eq:Phi2_main}), together with ${}_2F_1(\alpha,\beta,\gamma;z=0)=1$
\cite{ABST} and Eq. (\ref{eq:long_times})  [also using
Eqs. (\ref{eq:PPmap1}) and (\ref{eq:PPmap2})] gives
   \bea\label{eq:P_eq_finite_size}
\rho_{\cal T}^{\rm eq}(y_{\cal T}) &=& \frac{1}{(L-N\Delta)^N}
  \frac{(N_L+N_R+1)!}{N_L!\,N_R!} \nonumber \\
 &&  \times
  \left(\frac{L}{2}+y_{\cal T}-\Delta(1/2+N_L) \right)^{N_L} \nonumber \\
&&   \times
  \left(\frac{L}{2}-y_{\cal T}-\Delta(1/2+N_R) \right)^{N_R}.
  \eea
where $\rho_{\cal T}^{\rm eq}(y_{\cal T})=\rho_{\cal T}(y_{\cal T},t\rightarrow \infty|y_{{\cal T},0})$. This equation agrees with the equilibrium tPDF given in
Eq. (\ref{eq:P_eq_first}) (for $\Delta=0$), as it should \cite{some_clarifications}.
The results above shows the consistency of our analysis, and illustrates
the ergodicity of the finite SFD system. Calculating the second moment of the equilibrium tPDF, ${\cal S}(t\rightarrow \infty)= {\cal S}_{\rm eq}$,  gives
\be \label{eq:MSD_eq}
 {\cal S}_{\rm eq}= \left(\frac{1}{4}\right)^{N_R+1}
 \left(\frac{L-N\Delta}{2}\right)^2
%
\frac{\Gamma(1/2) \Gamma(2(N_R+1))}{\Gamma (N_R+1)\Gamma(N_R + 5/2) } ,
\ee
for $N_L=N_R$ where $\Gamma (z)$ is the the gamma function.  For $N\gg 1$ we can simplify the expression for the equilibrium tPDF as well as ${\cal S}_{\rm eq}$. If  a symmetric file is assumed, i.e. $N_L=N_R$ and ${\cal T} = (N+1)/2$, an asymptotic expansion of $\rho_{\cal T}^{\rm eq}(y_{\cal T})$ (using Stirling's approximation \cite{ABST}) gives the Gaussian PDF
\bea
\rho_{\cal T}^{\rm eq}(y_{\cal T}) &\approx&
\sqrt{\frac{2N}{\pi} \frac{1}{(L-N\Delta)^2}} \nonumber\\
&& \times
  \exp \left(-2N
	\left(
		\frac{y_{\cal T} }{L-N\Delta}
	\right)^2
\right),
\eea
from which we read off that
\be \label{eq:MSD_eq2}
 {\cal S}_{\rm eq} \approx \frac{(L-N\Delta)^2}{4N}.
\ee
Equation (\ref{eq:MSD_eq2}) can also be found directly from a large $N$ expansion of Eq. (\ref{eq:MSD_eq}) using Stirling's formula and assuming $N\approx 2N_R$. We point out that $\rho_{\cal T}^{\rm eq}(y_{\cal T}) \rightarrow 0$ for $N,L\rightarrow \infty $ even if the concentration $\varrho= N/L$ is kept fixed. This is consistent with the long time limit of Eq. (\ref{eq:P_bar_SFD}) which indeed goes to zero for large times. Finally, the analysis in this subsection also gives an estimate on the time required  to reach equilibrium, namely $t\gg\tau_{\rm eq}$.


\section{Conclusions and outlook} \label{sec:Conclusions}

In this study we have solved exactly a non-equilibrium statistical
mechanics problem: diffusion of $N$ hard-core interacting particles of
size $\Delta$ which are unable to pass each other in a one-dimensional
system of length $L$ with reflecting boundaries. In particular, we
obtained an exact expression for the probability density function $
\rho_{\cal T}(y_{\cal T},t|y_{{\cal T},0})$ (denoted tPDF) that a
tagged particle particle ${\cal T}$ is at position $y_{\cal T}$ at
time $t$ given that it at time $t=0$ was at position $y_{{\cal T},0}$.
We derived the tPDF by first finding the $N$-particle probability
density function ($N$PDF) via the Bethe ansatz, and then integrating
out the coordinates and taking the average over the initial positions
of all particles except one. The exact expression for $\rho_{\cal
  T}(y_{\cal T},t|y_{{\cal T},0})$ is found in Eqs, (\ref{eq:PPmap1}),
(\ref{eq:PPmap2}), (\ref{eq:P_bar3}) and (\ref{eq:Phi_main}), and
constitutes the main result of the paper.
For a large number of particles and for a tagged particle located  
roughly in the middle of the system, an asymptotic expansion of the  
tPDF  was derived [see Eq. (\ref{eq:P_bar_Bessel})]. Based on this  
equation, we found three time regimes of interest:
$(A)$ For short times, i.e. times much smaller than the collision time
$t\ll \tau_{\rm coll}=1/(\varrho^2D)$, where $\varrho=N/L$ is the
particle number concentration, the tPDF coincides with the Gaussian
probability density function which characterizes a free particle
[Eq. (\ref{eq:P_bar_t0})].
$(B)$ For intermediate times, $t\gg \tau_{\rm coll}$ but much smaller than  
the equilibrium time $t\ll \tau_{\rm
eq}=L^2/D$, a sub-diffusive single-file regime was found in which the  
tPDF is a Gaussian with an associated MSD proportional to $t^{1/2}$  
[Eq. (\ref{eq:P_bar_SFD})]. 
$(C)$ For times exceeding the equilibrium time $t\gg \tau_{\rm eq}$, the  
tPDF approaches a probability density of polynomial-type [Eq.~(\ref{eq:P_eq_finite_size})].

We point out that the sub-diffusive behavior for a tagged particle in time
regime $(B)$ is of fractional brownian motion type
\cite{goncalves,Lutz,our_FLE} rather than that of continuous-time random walks
(CTRWs) characterized by heavy-tailed waiting time densities
\cite{Klafter_87,MEKL1,MEKL2}. For such CTRW processes the probability density
function is not a Gaussian as for the current system. Further comparisons
between sub-diffusion in single-file systems and that occurring in CTRW theory
was pursued numerically in \cite{Marchesoni_06}.

The Bethe ansatz is often employed in quantum mechanics when many-body
systems are studied (e.g. quantum spin chains)
\cite{Batchelor_07,Thacker_81}, and also for stochastic many-particle
lattice problems \cite{Golinelli_06,SC}. We hope that the theoretical
analysis based on the Bethe ansatz presented here will stimulate
further progress in the field of single-file diffusion and that of
interacting random walkers. For instance, it would be interesting to
see whether our analysis could be extended to derive exact results
also for particles interacting through potentials other than hard-core
type \cite{semihardcore}, and for particles having different diffusion
constants \cite{diff_D}.

From the applied point of view, our exact expression for the tPDF
covers all time regimes and is straightforward to implement for
numerical computations. Therefore, we believe that our explicit
formula for $\rho_{\cal T}(y_{\cal T},t|y_{{\cal T},0})$, as well as
the approximate results for regimes $(A)-(C)$, will be useful for
experimentalists (see for instance \cite{Lin_05}) seeking to extract
system parameters such as particle size $\Delta$, system
size $L$, the particle's diffusion constants ($D$), and the number
($N_L$ and $N_R$) of particles to the left and right of the tagged
particle.

We finally note that the use of fluorescently labeled (tagged)
particles is of much use for studying biological systems.  The
understanding how the motion of such particles correlate with its
environment is therefore the key for grasping the behavior of such
systems in a quantitative fashion.


\section{Acknowledgments}

We are grateful to Bob Silbey, Mehran Kardar, Eli Barkai, Ophir
Flomenbom and Michael Lomholt for valuable discussions. L. L
acknowledges support from the Danish National Research Foundation, and
T. A. from the Knut and Alice Wallenberg Foundation.


 \appendix


\section{The Bethe ansatz}\label{sec:Bethe_ansatz_proofs}

In this section we show that the Bethe ansatz, Eq.
(\ref{eq:BetheAnsatz}), is a solution to the problem defined by Eqs.
(\ref{eq:DiffEq})-(\ref{eq:InitCond}). First, it is demonstrated that
Eq. (\ref{eq:BetheAnsatz}) satisfies the boundary conditions at the
ends of the box. Second, we show that the requirement that the
particles are unable to pass each other is satisfied by setting the
scattering coefficients to unity. Finally, it is demonstrated that 
Eq. (\ref{eq:BetheAnsatz}) also satisfies the initial condition.

\subsection{Boundary conditions at the ends of the box}

In this subsection it is proven that Eq. (\ref{eq:BetheAnsatz})
satisfies the the reflecting boundary conditions
Eq. (\ref{eq:NofluxCond1}) and (\ref{eq:NofluxCond2}) at $\pm\ell/2$
with an appropriate choice for $\phi(k_j,x_{j,0})$ [Eq.
(\ref{eq:I0_refl})]. The scattering coefficients are
set to $S_{ij}=1$ which is proven to be correct in the following
subsection. First, we define the function
 \be \label{eq:II}
        \lambda(k,z)=\phi(k,z)e^{-ik\ell/2} 
        = 2\cos[k(z+\ell/2)]\sum_{m=-\infty}^\infty e^{-2ikm\ell}  
\ee
which has the symmetry relation
\be\label{eq:II_symm}
        \lambda(k,z)=\lambda(-k,z).
  \ee
By taking the derivative of Eq. (\ref{eq:BetheAnsatz}) with respect to $x_1$, and evaluating at the left boundary $x_1=-\ell/2$ gives
\bea
&&\left.\frac{\partial P (\vec{x},t|\vec{x}_0)}{\partial x_1}\right|_{x_1=-\ell/2} 
= \nonumber\\ &&\int_{-\infty}^\infty  \frac{dk_1}{2\pi}  \cdots \int_{-\infty}^\infty \frac{dk_N}{2\pi}
 e^{-D(k_1^2+\ldots+k_N^2) t}  \Pi_{j=1}^N \phi (k_j,x_{j,0}) \nonumber\\
&& \times \Big[ ik_1 e^{ik_1x_1} \left(
                e^{ik_2x_2+\ldots+ik_Nx_N} +  {\rm perm. \ of \ } \{ k_2,...,k_N \} 
            \right)  \nonumber\\
&& \ \ \vdots \nonumber \\
&& \ \ \  + \, ik_i e^{ik_ix_1}  \left(
     e^{ik_2x_2+\ldots+ik_1x_i+\ldots+ik_N x_N} 
       \right.  \nonumber \\
&& \ \ \ + \left. {\rm perm. \ of \ } \{ k_2,...,k_{i-1},k_{i+1},...,k_N \} \right) \nonumber \\
&& \ \ \vdots \nonumber \\
&&Ê\ \ \,  + ik_N e^{ik_Nx_1} \left(e^{ik_2x_2+\ldots+ik_1x_N} \right.  
\nonumber \\
&& \ \ \ 
+ \left. {\rm perm. \ of \ } \{ k_1,...,k_{N-1} \} \right)\Big]_{x_1=-\ell/2}\nonumber\\
&&\nonumber\\
&&= \int_{-\infty}^\infty \frac{dk_2}{2\pi} \cdots \int_{-\infty}^\infty \frac{dk_N}{2\pi}
e^{-D(k_2^2+\ldots+k_N^2) t}\Pi_{j=2}^N \phi (k_j,x_{j,0})\nonumber\\
& &\times \int_{-\infty}^\infty \frac{dk_1}{2\pi}(ik_1)e^{-Dk_1^2t}\lambda(k_1,x_{1,0}) + \nonumber \\
&& \ \ \vdots \nonumber \\
&&+ \int_{-\infty}^\infty \frac{dk_1}{2\pi} \cdots\int_{-\infty}^\infty \frac{dk_{i-1}}{2\pi}\int_{-\infty}^\infty \frac{dk_{i+1}}{2\pi}\cdots \int_{-\infty}^\infty \frac{dk_N}{2\pi}\nonumber\\
&&\times e^{-D(k_1^2+\ldots+k_{i-1}^2+k_{i+1}^2+\ldots+k_N^2) t}\Pi_{j=1,j\neq i}^N \phi (k_j,x_{j,0})\nonumber\\
&&\times \int_{-\infty}^\infty \frac{dk_i}{2\pi}(ik_i)e^{-Dk_i^2t}\lambda(k_i,x_{i,0}) +  \nonumber \\
&& \ \ \vdots \nonumber \\
&&+  \int_{-\infty}^\infty \frac{dk_1}{2\pi} \cdots \int_{-\infty}^\infty \frac{dk_{N-1}}{2\pi}
e^{-D(k_1^2+\ldots k_{N-1}^2) t}\nonumber\\
& &\times \, \Pi_{j=1}^{N-1} \phi (k_j,x_{j,0})
\int_{-\infty}^\infty \frac{dk_N}{2\pi}(ik_N)e^{-Dk_N^2t}\lambda(k_N,x_{N,0}) \nonumber\\
&&= 0, 
\eea
where 
$\int_{-\infty}^\infty dk_i k_i e^{-Dk_i^2t}\lambda(k_i,x_{i,0})=0$ 
[odd integrand, see Eq. (\ref{eq:II_symm})] was used in the last step.
Note that this calculation does not rely on any specific form of
$\phi(k_j,x_{j,0})$. It is only required that the symmetry relation
(\ref{eq:II_symm}) holds.
Furthermore, the dispersion relation $E(\vec{k}) = D(k_1^2+\ldots+k_N^2)$
was also used in the above derivation. However, it would work equally
well for any dispersion relation as long as $E(\vec{k}) = \sum_i
E(k_i)$ with $E(k_i)=E(-k_i)$ is valid.

A similar analysis as just presented shows that the Bethe ansatz
solution also satisfies the reflecting condition at $+\ell/2$ [Eq.
(\ref{eq:NofluxCond2})]. In fact, since the Bethe ansatz is invariant
under the coordinate transformation $x_i\leftrightarrow x_j$, gives
$[\partial P(\vec{x},t|\vec{x}_0)/ \partial x_j ]_{x_j=\pm \ell/2}=0$
for {\em all} $x_j$.

\subsection{Single-file condition: particles are unable to overtake}

In this subsection it is shown that the condition that the particles
are unable to pass each other, Eq. (\ref{eq:NopassCond}), is satisfied
for scattering coefficients given by $S_{ij}=1$ in the Bethe ansatz
solution Eq. (\ref{eq:BetheAnsatz}). We start off by expressing ${\cal
  P}(\vec{x},t|\vec{x}_0)$ in two alternative ways:
%
\begin{widetext}
\bea\label{eq:BetheAnsatz_Sij_proof1}
{\cal P}(\vec{x},t|\vec{x}_0) &=& 
 \int_{-\infty}^\infty  \frac{dk_1}{2\pi}
 \int_{-\infty}^\infty  \frac{dk_2}{2\pi} \cdots 
 \int_{-\infty}^\infty \frac{dk_N}{2\pi}
\, e^{-D(k_1^2+\ldots+k_N^2) t} 
\Pi_{j=1}^N \phi (k_j,x_{j,0}) \nonumber\\
&&\times \Big[ 
e^{ik_1x_j} 
   \left(
  e^{ik_jx_1+\ldots+ik_{j-1}x_{j-1}+ik_{j+1}x_{j+1}+ik_{j+2}x_{j+2}
  +\ldots+ik_Nx_N} + {\ \rm all \ other \ permut. \ of \ } 
  \{ k_2,...,k_N \} \right) \nonumber\\
&& +\, e^{ik_2x_j}
   \left( e^{ik_1x_1+ik_jx_2+\ldots +ik_{j-1}x_{j-1}+ik_{j+1} 
   x_{j+1}+ik_{j+2}x_{j+2}+\ldots+ik_Nx_N}
  + { \ \rm all \ other \ permut. \ of \ } \{ k_1,k_3,...,k_N\}
  \right)  \nonumber\\
&& \vdots \nonumber \\
&&  +\, e^{ik_Nx_j}
        \left(e^{ik_1x_1+\ldots+ik_{j-1}x_{j-1}+ik_{j+1}x_{j+1}+ik_{j+2}
        x_{j+2}+\ldots+ik_jx_N} + {\ \rm all \ other \ permut. \ of \ } 
        \{ k_1,...,k_{N-1}\} \right)
 \Big]
\eea  
and
\bea\label{eq:BetheAnsatz_Sij_proof2}
{\cal P}(\vec{x},t|\vec{x}_0) &=& 
\int_{-\infty}^\infty  \frac{dk_1}{2\pi}  \int_{-\infty}^\infty \frac{dk_2}{2\pi} \cdots \int_{-\infty}^\infty \frac{dk_N}{2\pi}e^{ -D(k_1^2+\ldots+k_N^2)t}\Pi_{j=1}^N \phi (k_j,x_{j,0}) \nonumber\\
&& \times \Big[ 
e^{ik_1x_{j+1}} 
 \left ( e^{ik_{j+1}x_1+\ldots+ik_{j-1}x_{j-1}+ik_jx_j+ik_{j+2}x_{j+2}
                 +\ldots+ik_Nx_N} + { \rm all \ other \ permut. \ of \ } 
                 \{ k_2,...,k_N\} \right) \nonumber\\
& & +\, e^{ik_2x_{j+1}} 
        \left( e^{ik_1x_1+ik_{j+1}x_2+\ldots +ik_{j-1}x_{j-1}+ik_jx_j+
        ik_{j+2}x_{j+2}+\ldots+ik_Nx_N} + { \rm all \ other
         \ permut. \ of \ } \{ k_1,k_3,...,k_N\} \right) \nonumber\\
&& \vdots \nonumber \\
&&  +\, e^{ik_Nx_{j+1}} 
        \left(e^{ik_1x_1+\ldots+ik_{j-1}x_{j-1}+ik_jx_j+ik_{j+2}x_{j+2}+
        \ldots+ik_{j+1}x_N} + { \rm all \ other \ permut. \ of \ }
         \{ k_1,...,k_{N-1}\} \right)
      \Big].
  \eea
Using the above equations one finds
 \bea
 &&\left( \left.
\frac{\partial {\cal P}(\vec{x},t|\vec{x}_0)}{\partial x_{j+1}} - \frac{\partial {\cal P}(\vec{x},t|\vec{x}_0)}{\partial x_{j}}
\right) \right|_{x_{j+1}=x_j} =
\int_{-\infty}^\infty  \frac{dk_1}{2\pi}  \int_{-\infty}^\infty \frac{dk_2}{2\pi} \cdots \int_{-\infty}^\infty \frac{dk_N}{2\pi}e^{-D(k_1^2,\ldots,k_N^2)t}\Pi_{j=1}^N \phi (k_j,x_{j,0}) \nonumber\\
& & \times \Big[ 
ik_1 e^{ik_1x_j} 
        \left( e^{ik_{j+1}x_1+\ldots+ik_{j-1}x_{j-1}+ik_j x_{j+1}+
        ik_{j+2}x_{j+2}+\ldots+ik_Nx_N}\right.\nonumber\\
&&      \left. - e^{ik_j x_1+\ldots+ik_{j-1}x_{j-1}+ik_{j+1}x_{j+1}+
        ik_{j+2}x_{j+2}+\ldots+ik_Nx_N}
    + { \rm all \ other \ permut. \ of \ } \{ k_2,...,k_N\}\right)
     \nonumber\\
&& + \, ik_2 e^{ik_2x_j}
 \left( e^{ik_1 x_1+ik_{j+1}x_2+\ldots +ik_{j-1}x_{j-1}+ik_j 
 x_{j+1}+ik_{j+2}x_{j+2}+\ldots+ik_Nx_N} \right. \nonumber\\
&& \left. - e^{ik_1x_1+ik_{j}x_2+\ldots +ik_{j-1}x_{j-1}+ik_{j+1}x_{j+1}+
 ik_{j+2}x_{j+2}+\ldots+ik_Nx_N}
 + { \rm all \ other \ permut. \ of \ } \{ k_1,k_3,...,k_N\} \right)
 \nonumber\\
&& \vdots \nonumber \\
& &  +\,  ik_N e^{ik_Nx_j} (e^{ik_1x_1+\ldots+ik_{j-1}x_{j-1}
         +ik_jx_{j+1}+ik_{j+2}x_{j+2}+\ldots+ik_{j+1}x_N}\nonumber\\
&&  \left. - e^{ik_1x_1+\ldots+ik_{j-1}x_{j-1}+ik_{j+1} 
 x_{j+1}+ik_{j+2}x_{j+2}+\ldots+ik_jx_N}
 + { \rm all \ other \ permut. \ of \ } \{ k_1,...,k_{N-1}\}\right) 
 \big]=0
  \eea
  \end{widetext}
%
where it was used that each parenthesis is identically zero due to
the cancellation of the $2(N-1)!$ terms after permutation over all
allowed momenta.  Note that this derivation is independent of the
choice of $\phi(k_j,x_{j,0})$ and $E(\vec{k})$.

\subsection{Initial condition}
In this subsection we show that the Bethe ansatz Eq.
(\ref{eq:BetheAnsatz}) agrees with the initial condition Eq.
(\ref{eq:InitCond}) when $t\rightarrow~0$. By defining
  \be
\Omega(x,y)=\int_{-\infty}^\infty \frac{dk_j}{2\pi} e^{ik_j x} \phi (k_j,y),
  \ee
Eq. (\ref{eq:BetheAnsatz}) reads
\bea \label{eq:P_small_t}
        {\cal P}(\vec{x},t\rightarrow 0|\vec{x}_0)&=&\Omega (x_1,x_{1,0})       \Omega (x_2,x_{2,0})\cdots \Omega (x_N,x_{N,0})\nonumber\\
        &\hspace{-1cm}+& \hspace{-0.5cm}\Omega (x_1,x_{2,0})\Omega      (x_2,x_{1,0})\cdots \Omega (x_N,x_{N,0})\nonumber\\
        &\hspace{-1cm}+& \hspace{-0.5cm}{ \rm all \ other \ permut. \ of \ }    \{ x_{1,0},...,x_{N,0}\}.
  \eea
Using the explicit expression for $\phi(k,y)$ found in
Eq. (\ref{eq:I0_refl}) and that $\delta(x-z)=(2\pi)^{-1}
\int_{-\infty}^\infty dk \, e^{ik(x-z)} $ leads to
\be
        \Omega(x,y)=
        \sum_{m=-\infty}^\infty \delta(x-y+2m\ell)
        + \delta(x+y+(2m+1)\ell). 
\ee
For all $m\neq  0$ the $\delta$-functions are non-zero for coordinates lying outside of ${\cal R}$ and ${\cal R}_0$ where, per definition, ${\cal P}(\vec{x},t|\vec{x}_0)\equiv 0$  (see discussion in section Sect. \ref{sec:ProblemDef}). For $m=0$, it is only the first $\delta$-function in the sum, $\delta(x-y)$, that contributes to the $N$PDF. Furthermore, since all terms except the first one in Eq. (\ref{eq:P_small_t}) are zero due to the fact that ${\cal P}(\vec{x},t|\vec{x}_0)=0$ outside ${\cal R}$, one obtains  
\be
{\cal P}(\vec{x},t\rightarrow 0|\vec{x}_0) = \delta(x-x_0)\cdots \delta(x_N-x_{N,0})
\ee
as $t\rightarrow 0$ which is the desired result.



\section{Laplace transform of $\psi(x_i,x_{j,0};t)$}\label{sec:eigenmode_exp}

In this section it is shown explicitly how one can go from the
one-particle PDF $\psi(x_i,x_{j,0};t)$ for a
particle in a box expressed in terms of Gaussians [Eq.
(\ref{eq:P_single_refl})], to the eigenmode expansion
[Eqs.~(\ref{eq:P_single_refl2})-(\ref{eq:cp_sp})] used in this
paper. First, the summation in Eq. (\ref{eq:P_single_refl}) is divided
schematically into
$\sum_{m=-\infty}^\infty f_m = f_{m=0}+
\sum_{m=1}^\infty [f_m + f_{-m}]$.
Then, using the Laplace transform
${\cal L}[(\pi t)^{-1/2}e^{-a/(4t)}] = s^{-1/2}e^{-a \sqrt{s}}$ ($a>0$) %
 \cite{ABST} one finds
\bea \label{eq:Psi_s}
&& \psi(x_i,x_{j,0};s) = {\cal Q}(s) + 
 \frac{e^{-(x_i+x_{j,0}+\ell)\sqrt{s/D}}}{\sqrt{4Ds}} \nonumber\\
&& \ \ \ + \Big[ \frac{e^{-(x_i-x_{j,0})\sqrt{s/D}} 
+e^{(x_i-x_{j,0})\sqrt{s/D}}]}{\sqrt{4Ds}} \nonumber\\ 
&& \ \ \ + \frac{e^{-(x_i+x_{j,0}+\ell)\sqrt{s/D}} +
 e^{(x_i+x_{j,0}+\ell)\sqrt{s/D}}}{\sqrt{4Ds}} \Big] \nonumber\\
&&\ \ \ \times 
\sum_{m=1}^\infty e^{-2m\ell \sqrt{s/D}} 
  \eea
where
%
%
%
%
\be
{\cal Q} (s)= \left\{
  \begin{array}{l}\label{eq:Q_s}  
  (4Ds)^{-1/2} e^{-(x_i-x_{j,0})\sqrt{s/D}}, \  x_i \ge x_{j,0}\\
  (4Ds)^{-1/2} e^{-(x_{j,0}-x_i)\sqrt{s/D}}, \  x_i \le x_{j,0}.
  \end{array} 
  \right. 
\ee
Considering the cases $x_i>x_{j,0}$ and $x_i<x_{j,0}$ separately, and that
$
\sum_{m=1}^\infty e^{-2m\ell \sqrt{s/D}} = (e^{2\ell \sqrt{s/D}}-1)^{-1}
$
, leads to
\bea\label{eq:Psi_s_main}
        &&\psi(x_i,x_{j,0};s) =\frac{1}{\sqrt{sD} \sinh[\ell\sqrt{s/D}]} 
        \nonumber \\
   &&\times \left\{
   \begin{array}{ll}
         \cosh[(\ell/2+x_i) \sqrt{s/D}]  \\ 
                 \ \ \times 
                 \cosh[(\ell/2-x_{j,0})\sqrt{s/D}],
             & x_i\le  x_{j0}\\
             \\
             \cosh[(\ell/2-x_i)\sqrt{s/D}] \\ 
                 \ \ \times 
             \cosh[(\ell / 2+x_{j,0})\sqrt{s/D}],
             & x_i\ge x_{j,0}
   \end{array} \right.
  \eea
which after elementary trigonometric manipulations result in Eq. (\ref{eq:Psi_s2}).


\section{Extended phase space integration}\label{sec:extended_int}
\begin{figure}
 \includegraphics[width= 0.8\columnwidth]{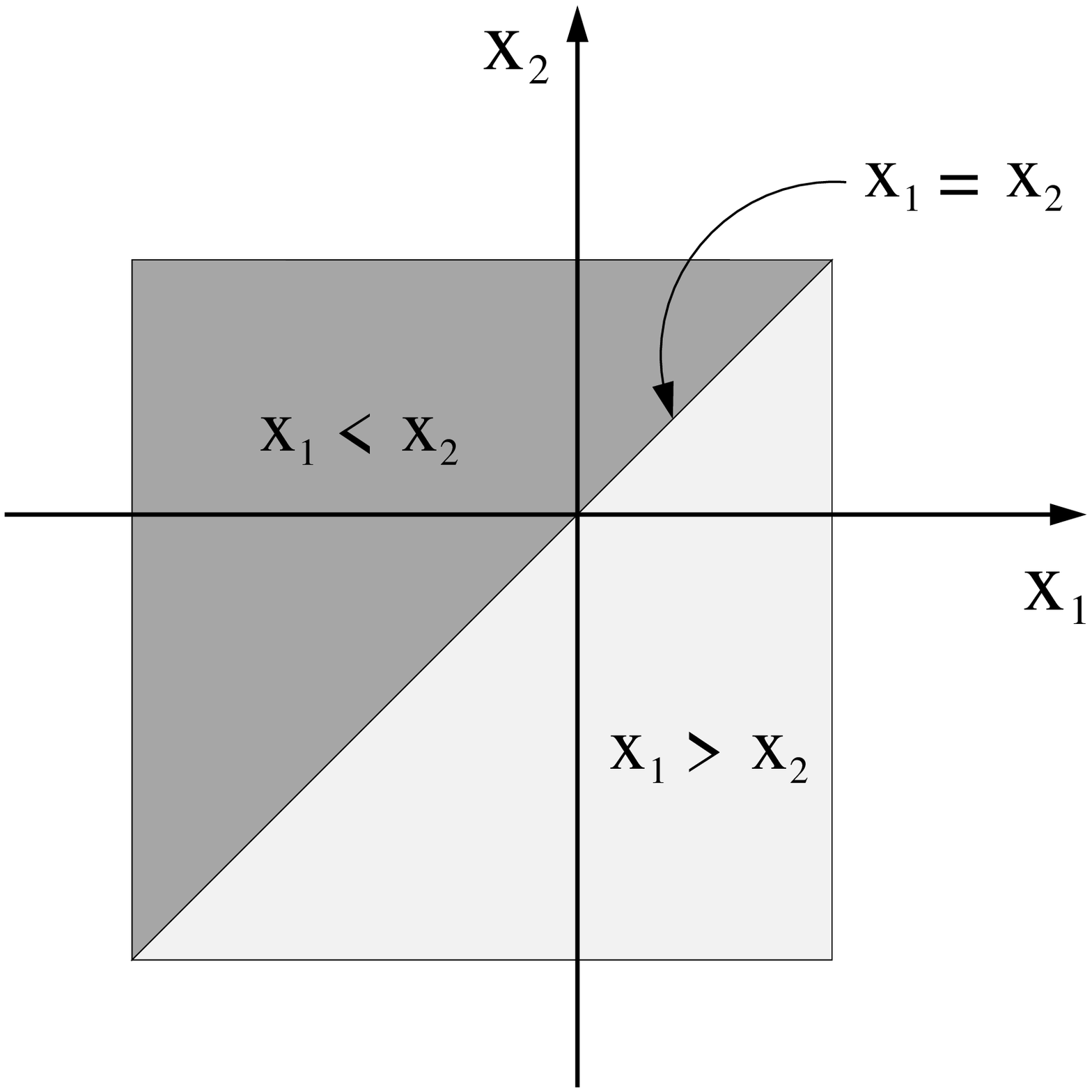}
 \caption{Integration area (the darker upper area, $x_1<x_2$) for the
integral defined in Eq. (\ref{eq:I3_1}). For any function which is
symmetric under $x_1\leftrightarrow x_2$ the integration over the
lower triangle ($x_1>x_2$) yields the same result as integration over
the same function over the upper triangle ($x_1<x_2$). One may
therefore extend the integration area to the full rectangle above
provided one divides the corresponding result by 2.}
 \label{fig:int_area}
\end{figure}
When the tPDF is integrated out from the $N$PDF we need to resolve the following type of integral:
\be\label{eq:Ik}
{\cal I} (x_{\cal T})=
\int_{ {\cal R} \backslash x_{\cal T} } dx_1' \cdots dx_{{\cal T}-1}'  
        dx_{{\cal T}+1}' \cdots dx_N'
{\cal P} (\vec{x}',t|\vec{x}_0)
\ee
over the region ${\cal R}$ [Eq. (\ref{eq:region_R})] with the tagged
particle coordinate $x_{\cal T}$ left out. As pointed previously in
the paper, the Bethe ansatz solution [Eq. (\ref{eq:BetheAnsatz})] is
symmetric under the transformation $x_i\leftrightarrow x_j$ when all
scattering coefficients are given by $S_{ij}=1$. This allows the
integration of ${\cal P}(\vec{x},t|\vec{x}_0)$ in Eq. (\ref{eq:Ik}) to
be extended to the whole hyperspace $x_j\in [-\ell/2, x_{\cal T}],
j=1,...,{\cal T}-1$ and $x_j\in [x_{\cal T}, \ell/2], j={\cal
  T}+1,...,N$. This is most easily demonstrated in an example which
then is extended to the general situation. Consider the case of three
particles where particle three is tagged ${\cal T}=3$:
\be\label{eq:I3_1}
        {\cal I} (x_3)=\int_{-\ell/2<x_1<x_2<x_3<\ell/2} dx_1 dx_2 {\cal P}     (x_1,x_2,x_3,t|\vec{x}_0).
\ee
The integration area in the $(x_1,x_2)$-plane is sketched in Fig. \ref{fig:int_area} (upper dark triangle). Since ${\cal P}(\vec{x},t|\vec{x}_0)$ is
invariant under $x_1\leftrightarrow x_2$, integration over the lower
triangle gives the same result, i.e.:
\be\label{eq:I3_2}
        {\cal I}(x_3) = \int_{-\ell/2<x_2<x_1<x_3<\ell/2}
        dx_1 dx_2 {\cal P}      (x_1,x_2,x_3,t|\vec{x}_0).
\ee
If Eqs. (\ref{eq:I3_1}) and (\ref{eq:I3_2})  are added, ${\cal I} (x_3)$ can be expressed as an integral over the full rectangle
  \be
{\cal I}(x_3)=\frac{1}{2}\int_{-\ell/2}^{x_{\cal T}} dx_1 \int_{-\ell/2}^{x_{\cal T}} dx_2 {\cal P}(x_1,x_2,x_3,t|\vec{x}_0)
  \ee
For the general case, the integration can be extended for any particle number to the left of the tagged particle. This means that it is possible to go from integration over the phase space
$-\ell/2<x_1<...<x_{{\cal T}-1}<x_{\cal T}$ 
to 
$-\ell/2<x_j<x_{\cal T}$ for $j=1,...,{\cal T}-1$  
provided that we divide by $N_L!$. This holds also for $N_R$ particles to the right and Eq. (\ref{eq:Ik}) can in general be written as
\bea
&&{\cal I} (x_{\cal T}) =  \frac{1}{N_L!N_R!}
\int_{-\ell/2}^{x_{\cal T}}dx_1\int_{-\ell/2}^{x_{\cal T}} dx_2 \cdots 
\\
&&\times \int_{-\ell/2}^{x_{\cal T}} dx_{{\cal T}-1} \int_{x_{\cal T}}^{\ell/2} dx_{{\cal T}+1}\cdots \int_{x_{\cal T}}^{\ell/2} dx_N {\cal P}(\vec{x},t|\vec{x}_0). \nonumber
\eea
Since ${\cal P}(\vec{x},t|\vec{x}_0)$ is also invariant under $x_{i,0}\leftrightarrow x_{j,0}$, a similar extended phase-space technique is valid for integrations over the initial particle positions ${\cal R}_0$.

\section{From the $N$PDF to the \lowercase{t}PDF}\label{sec:PDF_integration}

In this section the tPDF is obtained from the integral (\ref{eq:OnepartDistr}) using the Bethe ansatz $N$PDF (\ref{eq:BetheAnsatz2}) explicitly. A similar combinatorial analysis to that found in \cite{RKH} gives

\bea\label{eq:P_bar2}
&&\rho(x_{\cal T},t|x_{{\cal T},0}) = \frac{1}{N_L!N_R!} \Big\{ \nonumber\\
&&\sum_{q=0}^{{\rm min}\{N_L,N_R\}}  H^0(q) (\psi_L^L)^{N_L-q}(\psi_R^L)^q \psi (\psi_L^R)^q (\psi_R^R)^{N_R-q}\nonumber\\
&& + \hspace{-0.5cm} \sum_{q=0}^{{\rm min}\{N_L-1,N_R\}} \hspace{-0.5cm} 
H_L^L(q) (\psi_L^L)^{N_L-q-1}(\psi_R^L)^q \psi^L\psi_L (\psi_L^R)^q (\psi_R^R)^{N_R-q}\nonumber\\
&& + \hspace{-0.5cm} \sum_{q=0}^{{\rm min}\{N_L,N_R-1\}}\hspace{-0.5cm}
 H_R^R(q) (\psi_L^L)^{N_L-q}(\psi_R^L)^q \psi^R \psi_R (\psi_L^R)^q (\psi_R^R)^{N_R-q-1}\nonumber\\
&& + \hspace{-0.5cm} \sum_{q=0}^{{\rm min}\{N_L-1,N_R-1\}} \hspace{-0.5cm}
H_L^R(q) (\psi_L^L)^{N_L-q-1}(\psi_R^L)^{q+1} \psi^R \psi_L \nonumber \\ 
&& \hspace{1.5cm} \times (\psi_L^R)^q (\psi_R^R)^{N_R-q-1}\nonumber\\ 
&&+ \hspace{-0.5cm} \sum_{q=0}^{{\rm min}\{N_L-1,N_R-1\}} \hspace{-0.5cm} H_R^L(q) (\psi_L^L)^{N_L-q-1}(\psi_R^L)^q \psi^L\psi_R
\nonumber \\ 
&& \hspace{1.5cm} \times (\psi_L^R)^{q+1} (\psi_R^R)^{N_R-q-1} \Big \}
  \eea
with the combinatorial factors
\bea \label{eq:H}
H^0(q)&=&{N_L \choose q}{N_R \choose q} N_L! N_R!, \nonumber\\
H_L^L(q)&=&{N_L-1 \choose q} {N_R \choose q} N_L! N_R! N_L, \nonumber\\ 
H_R^R(q)&=&{N_L \choose q} {N_R-1 \choose q} N_L! N_R! N_R, \nonumber\\
H_L^R(q)&=&{N_L-1 \choose q} {N_R \choose q+1} N_L! N_R! N_L,\nonumber\\
H_R^L(q)&=&{N_L \choose q+1} {N_R-1 \choose q} N_L! N_R! N_R,
\eea
and integrals (leaving arguments $x_{{\cal T},0}$ and $x_{\cal T}$ implicit)
\bea\label{eq:integrals_def}
\psi_L^L(t)&=&f_L\int_{-\ell/2}^{x_{\cal T}}dx_i  \int_{-\ell/2}^{x_{{\cal T},0}}dx_{j,0} \psi(x_i,x_{j,0};t) \nonumber\\
\psi_R^R(t)&=&f_R  \int_{x_{\cal T}}^{\ell/2}dx_i  \int_{x_{{\cal T},0}}^{\ell/2} dx_{j,0} \psi(x_i,x_{j,0};t) \nonumber\\
\psi_L^R(t)&=&f_L \int_{x_{\cal T}}^{\ell/2}dx_i \int_{-\ell/2}^{x_{{\cal T},0}} dx_{j,0}  \psi(x_i,x_{j,0};t) \nonumber\\
\psi_R^L(t)&=&f_R  \int_{-\ell/2}^{x_{\cal T}}dx_i \int_{x_{{\cal T},0}}^{\ell/2} dx_{j,0}   \psi(x_i,x_{j,0};t) \nonumber\\
\psi^L(t)&=&\int_{-\ell/2}^{x_{\cal T}} dx_i \psi(x_i,x_{j,0};t) \nonumber\\
\psi^R(t)&=&\int_{x_{\cal T}}^{\ell/2}dx_i \psi(x_i,x_{j,0},t) \nonumber\\
\psi_L(t)&=&f_L \int_{-\ell/2}^{x_{{\cal T},0}}dx_{j,0} \psi(x_j,x_{j,0};t) \nonumber\\
\psi_R(t)&=&f_R\int_{x_{{\cal T},0}}^{\ell/2}dx_{j,0}  \psi(x_j,x_{j,0};t) 
\eea
The prefactors $f_L$ and $f_R$ are found in Eq. (\ref{eq:fL_fR}), and
correspond to uniform distributions to the left and right of the
tagged particle according to which the surrounding particles are
initially placed.  They appear when integrals over initial coordinates
are performed. Also, it is easy to see from normalization that
$\psi^L(t)+\psi^R(t)=1$, $\psi_L^L(t)+\psi_L^R(t)=1$ and
$\psi_R^R(t)+\psi_R^L(t)=1$.


If considering a single particle in a box of length $\ell$, the  integrals defined in Eq. (\ref{eq:integrals_def}) are easily interpreted as follows.  First, $\psi_L^L$ ($\psi_R^R$) is the probability that single particle is to the left (right) of $x_{\cal T}$ at time $t$ given that it started, with an equal probability, anywhere to the left (right) of $x_{{\cal T},0}$. Similar interpretations hold also for $\psi_L^R$ and $\psi_R^L$. The quantity $\psi_L$ ($\psi_R$) is the probability that a single particle is at position $x_{\cal T}$ given that the particle started somewhere to the left (right) of $x_{{\cal T},0}$. Finally, $\psi^L$ ($\psi^R$) is the probability that a single particle is to the left (right) of $x_{\cal T}$ at time $t$ given that it started at position~$x_{{\cal T},0}$.



\section{Integrals of $\psi(x_i,x_{j,0};s)$ in Laplace-space}\label{sec:Laplace_integrals}

In this section, exact expressions as well as limiting forms for the
integrals appearing in Eq. (\ref{eq:integrals_def}) are given in the
Laplace-domain. Using the Laplace-transformed one-particle PDF
$\psi(x_i,x_{j,0};s)$ found in Eq. (\ref{eq:Psi_s_main}), the
integrals defined in Eq. (\ref{eq:integrals_def}) (with time $t$
replaced by Laplace variable $s$, and $x_{\cal T}$ and $x_{{\cal
    T},0}$ left out) are given~by
  \bea\label{eq:Psi_LL_s}
        &&\psi^L(s) 
         =\frac{1}{s \sinh[\ell\sqrt{s/D}]} 
        \nonumber \\
   &&\ \ \times \left\{
   \begin{array}{l}
         \sinh[(\ell/2+x_{\cal T}) \sqrt{s/D}]  \\ 
                 \ \ \times 
                 \cosh[(\ell/2-x_{{\cal T},0})\sqrt{s/D}],
             \ x_{\cal T}\le  x_{{\cal T},0}\\
             \\
             \sinh[\ell\sqrt{s/D}]- \sinh[(\ell/2-x_{\cal T})\sqrt{s/D}] \\ 
                 \ \ \times 
             \cosh[(\ell/ 2+x_{{\cal T},0})\sqrt{s/D}],
             \ x_{\cal T}\ge x_{{\cal T},0}
   \end{array} \right.
  \eea
  \bea
    &&\psi_L(s) 
         =\frac{f_L}{s \sinh[\ell\sqrt{s/D}]} 
        \nonumber \\
   &&\ \ \times \left\{
   \begin{array}{l}
         \sinh[\ell\sqrt{s/D}]-\cosh[(\ell/2+x_{\cal T}) \sqrt{s/D}]  \\ 
                 \ \ \times 
                 \sinh[(\ell/2-x_{{\cal T},0})\sqrt{s/D}],
             \ \ x_{\cal T}\le x_{{\cal T},0}\\
             \\
             \cosh[(\ell/2-x_{\cal T})\sqrt{s/D}] \\ 
                 \ \ \times 
             \sinh[(\ell/ 2+x_{{\cal T},0})\sqrt{s/D}],
             \ \ x_{\cal T}\ge x_{{\cal T},0}
   \end{array} \right.
%
\eea
\bea
  &&\psi_R(s) 
         =\frac{f_R}{s \sinh[\ell\sqrt{s/D}]} 
        \nonumber \\
   &&\ \ \times \left\{
   \begin{array}{l}
         \cosh[(\ell/2+x_{\cal T}) \sqrt{s/D}]  \\ 
                 \ \ \times 
                 \sinh[(\ell/2-x_{{\cal T},0})\sqrt{s/D}],
             \ \ x_{\cal T}\le  x_{{\cal T},0}\\
             \\
             \sinh[\ell\sqrt{s/D}] - \cosh[(\ell/2-x_{\cal T})\sqrt{s/D}] \\ 
                 \ \ \times 
             \sinh[(\ell/ 2+x_{{\cal T},0})\sqrt{s/D}],
             \ \ x_{\cal T}\ge x_{{\cal T},0}
   \end{array} \right.
  \eea
 \bea
   &&\psi_L^L(s) 
         =\frac{1}{s \sinh[\ell\sqrt{s/D}]} 
        \nonumber \\
   &&\ \ \times \left\{
   \begin{array}{l}
         \left(1-(x_{{\cal T},0}-x_{\cal T})f_L\right) \sinh[\ell\sqrt{s/D}] \\
           - f_L\sqrt{\frac{D}{s}}\sinh[(\ell/2+x_{\cal T}) \sqrt{s/D}]  \\ 
                 \ \ \times 
                 \sinh[(\ell/2-x_{{\cal T},0})\sqrt{s/D}],
             \ \ x_{\cal T}\le  x_{{\cal T},0}\\
             \\
             \sinh[\ell\sqrt{s/D}]\\
             - f_L\sqrt{\frac{D}{s}} 
             \sinh[(\ell/2-x_{\cal T})\sqrt{s/D}] \\ 
                 \ \ \times 
             \sinh[(\ell/ 2+x_{{\cal T},0})\sqrt{s/D}],
             \ \ x_{\cal T}\ge x_{{\cal T},0}
   \end{array} \right.
  \eea
  \bea
     &&\psi_R^R(s) 
         =\frac{1}{s \sinh[\ell\sqrt{s/D}]} 
        \nonumber \\
   &&\ \ \times \left\{
   \begin{array}{l}
                     \sinh[\ell\sqrt{s/D}]\\
             - f_R\sqrt{\frac{D}{s}} 
             \sinh[(\ell/2+x_{\cal T})\sqrt{s/D}] \\ 
                 \ \ \times 
             \sinh[(\ell/ 2-x_{{\cal T},0})\sqrt{s/D}],
             \ \ x_{\cal T}\le x_{{\cal T},0}\\
             \\
              \left(1-(x_{\cal T}-x_{{\cal T},0})f_R\right) \sinh[\ell\sqrt{s/D}] \\
           - f_R\sqrt{\frac{D}{s}}\sinh[(\ell/2-x_{\cal T}) \sqrt{s/D}]  \\ 
                 \ \ \times 
                 \sinh[(\ell/2+x_{{\cal T},0})\sqrt{s/D}],
             \ \ x_{\cal T}\ge  x_{{\cal T},0} .
   \end{array} \right.
  \eea
The remaining three integrals follow from the
normalization conditions (arguments left implicit):
\be\label{eq:normalization_s}
        \psi^L+\psi^R=\frac{1}{s}, \ \
        \psi_L^L+\psi_L^R=\frac{1}{s},\ \
        \psi_R^R+\psi_R^L=\frac{1}{s}
\ee
Laplace inversion of the above relationships, using e.g. residue calculus \cite{Wunsch}, gives Eq. (\ref{eq:Psi_LL_etc}). In the following subsections we give asymptotic results in the (1) long and (2) short time limit for the expressions above.

\subsection{Long-time behavior}

The long-time behavior of Eqs.~(\ref{eq:Psi_LL_s})-(\ref{eq:normalization_s}) is obtained from a series expansion for $\ell\sqrt{s/D} \ll 1$, and reads (arguments left implicit)
\bea\label{eq:long_times}
&&\psi=   \psi_L=  \psi_R=\frac{1}{s\ell}, \ \  
\psi^L =  \psi_L^L = \psi_R^L=
     \frac{1}{s} \left(\frac{1}{2}+\frac{x_{\cal T}}{\ell}\right),
     \nonumber\\
&&\psi^R =  \psi_R^R =  \psi_L^R =
     \frac{1}{s}\left(\frac{1}{2}-\frac{x_{\cal T}}{\ell}\right),
\eea
The inverse transforms are found from $ {\cal L}^{-1}(s^{-1})=1$~\cite{ABST}. 

\subsection{Short-time behavior}

Short times is defined  here as 
$(\ell\pm x_{\cal T})\sqrt{s/D} ,(\ell\pm x_{{\cal T},0})\sqrt{s/D}\gg 1$,
i.e. times shorter than the time it takes to diffuse across the entire box.
The short time behavior of  Eqs. (\ref{eq:Psi_LL_s})-(\ref{eq:normalization_s}) is given by 
 \be
        \psi^L(s) \approx
        \left\{ \begin{array}{ll}
                \frac{1}{2s} e^{-(x_{{\cal T},0}-x_{\cal T})\sqrt{s/D}},
                &x_{\cal T}\le x_{{\cal T},0}\\
                \frac{1}{s} \left(1-\frac{1}{2}e^{-(x_{\cal T}-x_{{\cal T},0})
                \sqrt{s/D}}\right),
                &x_{\cal T}\ge x_{{\cal T},0}
        \end{array}\right.
\ee
\be\label{eq:Psi_L_s}
        \psi_L (s) \approx 
        \left\{ \begin{array}{ll}
                \frac{f_L}{s} \left(1-\frac{1}{2} e^{-(x_{{\cal T},0}-x_{\cal T})               \sqrt{s/D}}\right), \  
                &x_{\cal T}\le x_{{\cal T},0}\\
                \frac{f_L}{2s} e^{-(x_{\cal T}-x_{{\cal T},0})\sqrt{s/D}},
                &x_{\cal T}\ge x_{{\cal T},0}
   
        \end{array}\right.
  \ee
 \be
        \psi_R(s) \approx 
        \left\{ \begin{array}{ll}
                \frac{f_R}{2s} e^{-(x_{{\cal T},0}-x_{\cal T}) \sqrt{s/D}},
                &x_{\cal T}\le x_{{\cal T},0}\\
                \frac{f_R}{s}\left( 1-\frac{1}{2} e^{-(x_{\cal T}-x_{{\cal T},0})
                \sqrt{s/D}} \right), 
                &x_{\cal T}\ge x_{{\cal T},0}
        
        \end{array}\right.
\ee
\be
        \psi_L^L(s) \approx
        \left\{ \begin{array}{ll} 
            \frac{1}{s}\Big(1-(x_{{\cal T},0}-x_{\cal T})f_L\\
        \ \ -\frac{f_L}{2}\sqrt{\frac{D}{s}} 
            e^{-(x_{{\cal T},0}-x_{\cal T})\sqrt{s/D}}\Big),
            &x_{\cal T}\le x_{{\cal T},0}\\
                \frac{1}{s} \left(1-\frac{f_L}{2} \sqrt{\frac{D}{s}} 
                 e^{-(x_{\cal T}-x_{{\cal T},0})\sqrt{s/D}}\right),
                &x_{\cal T}\ge x_{{\cal T},0}
       
        \end{array}\right.
\ee
\be\label{eq:Psi_RR_s}
        \psi_R^R(s) \approx
                \left\{ \begin{array}{ll}  
                        \frac{1}{s}\left(1-\frac{f_R}{2}\sqrt{\frac{D}{s}} 
                        e^{-(x_{{\cal T},0}-x_{\cal T})\sqrt{s/D}}\right),
                        &x_{\cal T}\le x_{{\cal T},0}\\
                        \frac{1}{s}\Big(1-(x_{\cal T}-x_{{\cal T},0})f_R \\
                        \ \ -\frac{f_R}{2} \sqrt{\frac{D}{s}} 
                        e^{-(x_{\cal T}-x_{{\cal T},0}) \sqrt{s/D}}\Big),
                        &x_{\cal T}\ge x_{{\cal T},0}
                       
        \end{array}\right.
\ee
The remaining integrals follow from Eq. (\ref{eq:normalization_s}). 

Equations (\ref{eq:Psi_L_s})-(\ref{eq:Psi_RR_s}) can be inverted exactly
into time domain. Using standard formulas \cite{ABST} and introducing
\be\label{eq:eta}
        \eta=\frac{x_{\cal T}-x_{{\cal T},0}}{\sqrt{4Dt}},
\ee
leads to
\bea\label{eq:Psi_LL_etc_short_time}
&&\psi_L(t)=\frac{f_L}{2}(1-{\rm erf} \eta),
\ \ \psi_R(t)=\frac{f_R}{2}(1+{\rm erf} \eta)\nonumber\\
&&\psi^L(t)=\frac{1}{2}(1+{\rm erf} \eta),
\ \ \psi^R(t)=\frac{1}{2}(1-{\rm erf} \eta)\nonumber\\
&&\psi_L^L(t)=1-\frac{f_L}{2}\sqrt{\frac{4Dt}{\pi}}
   \left[e^{-\eta^2}+\sqrt{\pi}\eta ({\rm erf} \eta -1)\right]\nonumber\\
&&\psi_R^R(t)=1-\frac{f_R}{2}\sqrt{\frac{4Dt}{\pi}}
        \left[e^{-\eta^2}+\sqrt{\pi}\eta ({\rm erf} \eta + 1)\right] \nonumber\\
\eea
where ${\rm erf} z$ is the error-function \cite{ABST}. 
The expressions above are valid for times such that $4Dt/(\ell/2\pm
x_{{\cal T},0})^2\ll 1$ and $4Dt/(\ell/2\pm x_{\cal T})^2\ll
1$. Expanding the result in
Eq. (\ref{eq:Psi_LL_etc_short_time}) for small $\eta$ and using the normalization conditions Eq. (\ref{eq:normalization_s}), yields
\bea\label{eq:Psi_LL_etc_eta_expansion}
        &&\psi_L(t)=\frac{f_L}{2}\left(1-\frac{2 \eta}{\sqrt{\pi}}\right), 
        \ \ \psi_R(t)=\frac{f_R}{2} \left(1+\frac{2 \eta}{\sqrt{\pi}} \right)
         \nonumber\\
        &&\psi^L(t)=\frac{1}{2} \left(1+\frac{2 \eta}{\sqrt{\pi}} \right),
        \ \ \psi^R(t)=\frac{1}{2} \left(1-\frac{2 \eta}{\sqrt{\pi}} \right)
        \nonumber\\
        &&\psi_L^R(t)=\frac{f_L}{2}\sqrt{\frac{4Dt}{\pi}} 
           \left(1-\sqrt{\pi}\eta \right),
        \nonumber\\
        && \psi_R^L(t)=\frac{f_R}{2}\sqrt{\frac{4Dt}{\pi}}
        \left(1+\sqrt{\pi}\eta \right).
\eea
and (arguments left implicit)
\be\label{eq:normalization_s}
        \psi^L+\psi^R=1, \ \
        \psi_L^L+\psi_L^R=1,\ \
        \psi_R^R+\psi_R^L=1.
\ee
The limit where $t\rightarrow 0$ limit ($s\rightarrow \infty$) is most conveniently found from Eqs. (\ref{eq:Psi_LL_s})-(\ref{eq:normalization_s}) which, in  combination with ${\cal L}^{-1}(s^{-1})=1$, reads
\be\label{eq:Psi_L_t0}
\psi^L(t\rightarrow 0)=
                \left\{ \begin{array}{ll}  
                0,              & x_{\cal T} \le x_{{\cal T},0}\nonumber\\
                1,              & x_{\cal T} \ge x_{{\cal T},0}
        \end{array}\right.
\ee
\be
\psi_L(t\rightarrow 0)=
                \left\{ \begin{array}{ll}  
                f_L,   &x_{\cal T} \le x_{{\cal T},0}\nonumber\\
                0,     &x_{\cal T} \ge x_{{\cal T},0}
        \end{array}\right.
\ee
\be
\psi_R(t\rightarrow 0)=
                \left\{ \begin{array}{ll}  
                0,                      &x_{\cal T} \le x_{{\cal T},0}\nonumber\\
                f_R,            &x_{\cal T} \ge x_{{\cal T},0}
        \end{array}\right.
\ee
\be
\psi_L^L(t\rightarrow 0)=
                \left\{ \begin{array}{ll}  
                1-(x_{{\cal T},0}-x_{\cal T})f_L,
                                & x_{\cal T} \le x_{{\cal T},0}\nonumber\\
                1,               & x_{\cal T} \ge x_{{\cal T},0}
        \end{array}\right.
\ee
\be
\psi_R^R(t\rightarrow 0)=
                \left\{ \begin{array}{ll}  
                1,      & x_{\cal T} \le x_{{\cal T},0}\nonumber\\
                1-(x_{\cal T}- x_{{\cal T},0})f_R, 
                           &x_{\cal T} \ge x_{{\cal T},0}
        \end{array}\right.
\ee
Again, the remaining integrals follow from the normalization condition (\ref{eq:normalization_s}).

\section{Macroscopic dynamics - the dynamic structure factor and 
center-of-mass motion}\label{sec:macroscopic}

Macroscopic quantities for a single-file system is the same as for a
system consisting of non-interacting particles \cite{HA,Kutner_81}. By
``macroscopic'' we here refer to any quantity which is invariant under
$x_i\rightarrow x_j$, i.e. any one which do not "notice" if two particle are interchanged in the system. In this Appendix we explicitly
evaluate two such macroscopic quantities: (1) the dynamic structure
factor and (2) the PDF for the center-of-mass coordinate. We demonstrate that indeed they agree with known results for non-interacting systems.

\subsection{Dynamic structure factor}

The dynamic structure factor is tractable via scattering experiments
(see e.g. Ref. \cite{berne}), and is widely used in condensed matter
physics and crystallography. Considering a set of noise-driven
stochastic trajectories $X_1(t),...,X_N(t)$, the dynamic structure
factor is \cite{doi}
\be\label{eq:S_Q_T}
        S(Q,t)=\frac{1}{N} \sum_{i,j} \langle e^{iQ(X_i(t)-X_{j,0})}\rangle,
\ee
where the brackets denote an average over different realizations of
the noise. In terms of the $N$PDF and the equilibrium $N$PDF (see Sec.
\ref{sec:ProblemDef}), it is given by
\bea
        S(Q,t)&=&\frac{1}{N} \int_{\cal R} dx_1\cdots dx_N \int_{{\cal R}_0}    dx_{1,0}\cdots dx_{N,0}\nonumber\\
        && \times \, \sum_{i,j} e^{iQ(x_i-x_{j,0})}{\cal P}(\vec{x},t|\vec{x}_0)
        {\cal P}^{\rm eq}(\vec{x}_0)
\eea
where we averaged over the (equilibrium) initial positions. Since the
integrand above is invariant under $x_i\leftrightarrow x_j$ we can
apply the technique explained in Appendix \ref{sec:extended_int} to
extend the integrations over coordinates as well as initial positions
to $[-\ell/2,\ell/2]$, which leads to
\bea
        S(Q,t)&=&\frac{1}{N}\frac{1}{N!}\frac{1}{\ell^N} 
        \int_{-\ell/2}^{\ell/2} dx_1 \int_{-\ell/2}^{\ell/2} dx_2 \cdots 
        \int_{-\ell/2}^{\ell/2} dx_N  \nonumber\\
        &&\times  \int_{-\ell/2}^{\ell/2} dx_{1,0} \int_{-\ell/2}^{\ell/2} dx_{2,0}     \cdots \int_{-\ell/2}^{\ell/2} dx_{N,0}\nonumber\\
        && \times \, \sum_{i,j}e^{iQ(x_i-x_{j,0})} {\cal P}(\vec{x},t|\vec{x}_0),
\eea
where also Eq. (\ref{eq:P_N_eq}) was used. Inserting 
explicitly the $N$PDF from Eq. (\ref{eq:BetheAnsatz2}) yields
  \bea\label{eq:S_Q_t2}
S(Q,t)&=&\frac{1}{N}\frac{1}{N!}\frac{1}{\ell^N} \int_{-\ell/2}^{\ell/2} dx_1 \int_{-\ell/2}^{\ell/2} dx_2 \cdots \int_{-\ell/2}^{\ell/2} dx_N  \nonumber\\
&&\times  \int_{-\ell/2}^{\ell/2} dx_{1,0} \int_{-\ell/2}^{\ell/2} dx_{2,0} \cdots \int_{-\ell/2}^{\ell/2} dx_{N,0}\nonumber\\
&& \left[ \left (\sum_{i=1}^N e^{iQ(x_i-x_{i,0})}+\sum_{i,j,i\neq j} e^{iQ(x_i-x_{j,0})}\right)\right.\nonumber\\
&& \times \psi(x_1,x_{1,0};t)\psi(x_2,x_{2,0};t)\cdots\psi(x_N,x_{N,0};t)\nonumber\\
&& + \ {\rm remaining} \ N!-1 \ {\rm  terms} \Big].
  \eea
Defining
  \bea\label{eq:C_Q_etc}
C(Q)&=&\frac{1}{\ell} \int_{-\ell/2}^{\ell/2} dx_i \int_{-\ell/2}^{\ell/2}
 dx_{j,0} \psi(x_i,x_{j,0};t)  e^{iQ(x_i-x_{j,0})}\nonumber\\
F(Q)&=&\frac{1}{\ell^2} \int_{-\ell/2}^{\ell/2} dx_i \int_{-\ell/2}^{\ell/2} 
dx_{j,0}\int_{-\ell/2}^{\ell/2} dx_k \int_{-\ell/2}^{\ell/2} dx_{l,0} \nonumber\\
&& \times \psi(x_i,x_{j,0};t) \psi(x_k,x_{l,0};t)  e^{iQ(x_i-x_{l,0})}
  \eea
and noticing that all the $N!$ terms corresponding to permutations of
the initial particle positions in Eq. (\ref{eq:S_Q_t2}) give the same
contribution, leads to
\be\label{eq:S_Q_t3}
        S(Q,t)=C(Q)C(0)^{N-1}+(N-1)F(Q)C(0)^{N-2}.
\ee
This result is identical to that of  non-interacting
particles (obtained e.g. by putting, $S_{ij}\equiv 0$, in the Bethe Ansatz
Eq. (\ref{eq:BetheAnsatz}) and allowing the particles to be in the
full phase space $x_j\in [-\ell/2,\ell/2]$). The expression above is simplified by noticing that $C(0)=1$, due to the normalization of $\psi(x_i,x_{j,0};t)$.

As an example, we consider the case $\ell\rightarrow\infty$ and $\Delta = 0$ where
$\psi(x_i,x_{j,0};t)=(4\pi D t)^{-1/2}\exp[-(x_i-x_{j,0})^2/(4Dt)]$ 
can be used. A straightforward calculation of Eq. (\ref{eq:C_Q_etc}) for this case gives
\bea
        C(Q)&=&e^{-DQ^2t} \\
        F(Q)&=&\frac{2\pi}{\ell}\delta(Q) e^{-DQ^2t},
\eea
where $\delta(z)= (2\pi)^{-1} \int_{-\infty}^\infty d\alpha \, 
e^{i\alpha z}$ was used. Inserting the results above into
Eq. (\ref{eq:S_Q_t3}) gives
\be
        S(Q,t)=e^{-DQ^2t}, \ \ Q\neq 0
\ee
which is in agreement with standard results for non-interacting particles in one
dimension (see e.g. Ref. \cite{clark}).

\subsection{Center-of-mass dynamics}\label{sec:CM}

The probability density function for the center-of-mass (CM) coordinate $X$ is given by
\bea\label{eq:P_CM}
        &&P(X,t|\vec{x}_0)= \nonumber\\
        &&\ \ \int_{\cal R} dx_1 dx_2 \cdots dx_N  
        \delta \left(X- \frac{x_1+x_2+...+x_N}{N}\right) \nonumber\\
        &&\ \ \ \times P(\vec{x},t|\vec{x}_0),
\eea
where the $N$PDF $P(\vec{x},t|\vec{x}_0)$ is given in Eq.
(\ref{eq:BetheAnsatz}), and the integration region ${\cal R}$ is
defined in Eq. (\ref{eq:region_R}). Since the integrand is
invariant under $x_i\leftrightarrow x_j$, it is possible to extend the
integration region to $x_j\in [-\ell/2,\ell/2]$ (provided we divide by
$N!$), as was done in previous subsection. Also, using the integral
representation of the $\delta$-function from previous subsection, Eq.
(\ref{eq:P_CM}) can be rewritten as
\bea\label{eq:P_CM2}
        && P(X,t|\vec{x}_0)=\frac{1}{N!}\int_{-\infty}^\infty \frac{dQ}{2\pi} 
        \int_{- \ell/2}^{\ell/2} dx_1\cdots \int_{-\ell/2}^{\ell/2} dx_N
        \nonumber\\
        &&\ \ \times \exp\left[iQ\left(X-\frac{x_1+x_2+...+x_N}{N}\right)\right]
        P(\vec{x},t|\vec{x}_0).\nonumber\\
\eea
Combining this equation with Eqs. (\ref{eq:BetheAnsatz}) and
(\ref{eq:disp_rel}), and defining $g(k_j,Q)=\int_{-\ell/2}^{\ell/2}
dx_i \, e^{ix_i (k_j-Q/N)}$ gives
 \bea
P(X,t|\vec{x}_0)&=&\int_{-\infty}^\infty \frac{dQ}{2\pi} e^{iQX} 
\int_{-\infty}^{\infty} \frac{dk_1}{2\pi}\cdots \int_{-\infty}^{\infty} \frac{dk_N}{2\pi} \nonumber\\
&&\times e^{-D(k_1^2+...+k_N^2)t} \phi(k_1,x_{1,0})\cdots \phi(k_N,x_{N,0})\nonumber\\
&& \times g(k_1,Q)\cdots g(k_N,Q).
  \eea
Integrating over $k_j$ leads to
\be\label{eq:P_CM3}
        P(X,t|\vec{x}_0)=\int_{-\infty}^\infty \frac{dQ}{2\pi}
        e^{iQX} h(Q,x_{1,0})\cdots h(Q,x_{N,0}),
  \ee
where
\bea\label{eq:h_Q}
        h(Q,x_{j0})&=&\int_{-\infty}^\infty \frac{dk_j}{2\pi} e^{-Dk_j^2 t}
        \phi (k_j,x_{j0})g(k_j,Q)\nonumber\\
        &=&\int_{-\ell/2}^{\ell/2} dx_i\int_{-\infty}^\infty \frac{dk_j}{2\pi} 
        e^{-Dk_j^2t}\nonumber\\
        &&\times \phi(k_j,x_{j,0})e^{ix_i(k_j-Q/N)}
\eea
The result in Eq. (\ref{eq:P_CM3}) is identical to that of non-interacting particles, as can be shown straightforwardly by setting
$S_{ij}\equiv 0$ in Eq. (\ref{eq:BetheAnsatz}) and not restricting the
particles to the phase space region ${\cal R}$.

As an example, Eq. (\ref{eq:P_CM3}) is evaluated for an infinite system
($\ell \rightarrow \infty$) and $\Delta = 0$ where $\phi(k_j,x_{j,0})=e^{-ik_jx_{j,0}}$. For this case Eq. (\ref{eq:h_Q}) becomes
  \be
h(Q,x_{j,0})=\exp\left( -\frac{Q^2 Dt}{N^2}-\frac{ix_{j,0}Q}{N}\right)
  \ee
which when inserted in Eq. (\ref{eq:P_CM3}) and integrated over $Q$
gives the well known Gaussian for the CM
  \be
P(X,t|\vec{x}_0)=\frac{1}{(4\pi D_{CM}t)^{1/2}} 
\exp\left[-\frac{(X-X^0_{CM})^2}{4D_{CM}t}\right]
  \ee
where  $X^0_{CM}=\left(\sum_{i=1}^N x_{i,0}\right)/N$ and $D_{CM}=D/N$ denote the CM initial position, and diffusion constant, respectively.


\newpage
\bibliographystyle{unsrt}
\bibliography{refs}

\begin{thebibliography}{10}
\bibitem{footnote3}
The Bethe ansatz solution is often written in terms of discrete wave vectors
  \cite{Golinelli_06}. Using the Poisson summation formula gives the identity
  $\sum_{m=-\infty}^\infty\, e^{2 i m k_j \ell} = (\pi/\ell)
  \sum_{m=-\infty}^\infty \delta(k_j+m\pi/\ell)$, from which the Bethe ansatz
  solution, Eqs. (\ref{eq:BetheAnsatz})-(\ref{eq:I0_refl}), can be rewritten in
  term of discrete wave vectors $m_j \pi/\ell$
  ($m_j=-\infty,...,-1,0,1,...,\infty$) if desired.
\bibitem{footnote1}
Note that our convention, with coordinates in the range $[-L/2,L/2]$, is
  different to Ref. \cite{RKH} where the left (right) end of the system is at
  $0$ ($L$).
\bibitem{rel_to_rodenbeck_pdf}
The expression for the tPDF, Eq. (\ref{eq:P_bar3}), is (for $\Delta=0$) related
  to Eq. (61) in \cite{RKH} via elementary recursion relations for the Jacobi
  polynomial \cite{ABST}.
\bibitem{xi_y}
The variable $\xi$ defined in Eq. (\ref{eq:xi}) is related to the variable $y$
  in Ref. \cite{RKH} according to $\xi=1/y$.
\bibitem{zeta_def}
We inverted the definition (see Ref. \cite{Elliott_71}) $\cosh 2 \zeta =
  (1+\xi)/(1-\xi)$ to arrive at Eq. (\ref{eq:zeta}).
\bibitem{some_details}
In more detail, in arriving at the SFD expression Eq. (\ref{eq:P_bar_SFD}) for
  $\rho_{\cal T}(y_{\cal T},t|y_{{\cal T},0})$ we assumed $1/N\ll 1$,
  $\delta_N/N\ll 1$, $(4Dt)/(\ell/2\pm x_{\cal T})^2\ll 1$, $(4Dt)/(\ell/2\pm
  x_{{\cal T},0})^2\ll 1$, $|\eta|\ll 1$, $x_{{\cal T},0}/\ell\ll 1$ and
  $|\delta_N (x_{\cal T}-x_{{\cal T},0})|/\ell\ll 1$.
\end{thebibliography}

\end{document}